\documentclass[structabstract]{aa}
\usepackage{graphicx}
\usepackage{subfig}                
\usepackage{longtable}             
\usepackage{txfonts}
\usepackage{natbib}
\usepackage{rotating}
\bibpunct{(}{)}{;}{a}{}{,} 

\def\N2H{N$_2$H$^+$}

\def\HII{H\,{\sc ii}}

\def\kms{km~s$^{-1}$}

\def\s{$\pm$}

\def\s12{\mbox{$S_{\rm 1.2mm}$}}

\def\la{\mathrel{\mathchoice {\vcenter{\offinterlineskip\halign{\hfil
$\displaystyle##$\hfil\cr<\cr\sim\cr}}}
{\vcenter{\offinterlineskip\halign{\hfil$\textstyle##$\hfil\cr
<\cr\sim\cr}}}
{\vcenter{\offinterlineskip\halign{\hfil$\scriptstyle##$\hfil\cr
<\cr\sim\cr}}}
{\vcenter{\offinterlineskip\halign{\hfil$\scriptscriptstyle##$\hfil\cr
<\cr\sim\cr}}}}}
\def\ga{\mathrel{\mathchoice {\vcenter{\offinterlineskip\halign{\hfil
$\displaystyle##$\hfil\cr>\cr\sim\cr}}}
{\vcenter{\offinterlineskip\halign{\hfil$\textstyle##$\hfil\cr
>\cr\sim\cr}}}
{\vcenter{\offinterlineskip\halign{\hfil$\scriptstyle##$\hfil\cr
>\cr\sim\cr}}}
{\vcenter{\offinterlineskip\halign{\hfil$\scriptscriptstyle##$\hfil\cr
>\cr\sim\cr}}}}}

\def\sec{\hbox{$^{\prime\prime}$}}

\newcommand{\beq}{\begin{equation}}
\newcommand{\eeq}{\end{equation}}
\newcommand{\bdi}{\begin{displaymath}}
\newcommand{\edi}{\end{displaymath}}

\begin{document}


\title{Discovery of weak 6.7-GHz CH$_3$OH masers in a sample of high-mass Hi-GAL sources }


\author{
L. Olmi \inst{\ref{inst1},\ref{inst2}} \and
E. D. Araya \inst{\ref{inst3}} \and
P. Hofner \inst{\ref{inst4}} \and
S. Molinari\inst{5} \and
J. Morales Ortiz \inst{\ref{inst2}} \and
L. Moscadelli \inst{\ref{inst1}} \and
M. Pestalozzi\inst{5}
}

\institute{
	  INAF, Osservatorio Astrofisico di Arcetri, Largo E. Fermi 5,
          I-50125 Firenze, Italy, \email{olmi.luca@gmail.com}                   \label{inst1}  
\and
	  University of Puerto Rico, Rio Piedras Campus, Physics Dept., Box 23343, 
	  UPR station, San Juan, Puerto Rico, USA                              \label{inst2} 
\and
          Physics Department, Western Illinois University, 1 University Circle, 
	  Macomb, IL 61455, USA                                                \label{inst3}
\and
 	  New Mexico Institute of Mining and Technology, Physics Department, 
          801 Leroy Place, Socorro, NM 87801, USA. Adjunct Astronomer at the National Radio Astronomy Observatory.  \label{inst4}
\and
	  Istituto di Fisica dello Spazio Interplanetario - INAF, via Fosso del 
	  Cavaliere 100, I-00133 Roma, Italy                                    \label{inst5}
          }

\date{Received; accepted }


\abstract
{Maser lines from different molecular species,
including water, hydroxyl, and methanol, are
common observational phenomena associated with
massive star forming regions.
In particular, since its discovery, the 6.7-GHz methanol maser has been
recognized as one of the clearest signposts to the formation of young
high-mass stars.
}
{The methanol maser thus appears as an ideal tool to study the early phases of massive star formation. 
However, it is difficult to establish the exact start of the methanol maser phase, and it would
then be interesting to detect and study low-flux density methanol masers 
(i.e., $\la 0.1\,$Jy or even $\ll 0.1\,$Jy), in order to determine if they can 
effectively be used to mark a specific evolutionary phase in high-mass star formation. 
}
{Past surveys have been unable to systematically detect many low-flux density methanol masers, 
and thus we do not yet know how many such masers exist in the Galaxy 
and what is their physical nature. 
A large sample of massive cores can now be found in the Herschel infrared GALactic 
Plane Survey (Hi-GAL), which we have used to search for methanol and excited OH masers towards
a sample of pre- and proto-stellar high-mass clumps using the Arecibo telescope.
}
{Out of a sample of 107 observed Hi-GAL sources we detected a total of 32 methanol masers, 
with 22 sources being new and weak (median peak flux density 0.07\,Jy) detections, 
in the Galactic longitude range $[32^{\circ}.0,59^{\circ}.8]$. 
We also detected 12 6.035-GHz OH maser, with 9 objects being new detections.
Our survey covers a similar range of source distances as the ``Arecibo Methanol Maser Galactic
Plane Survey'' (or AMGPS, \citealp{pandian2007a}), but  the methanol masers detected by us 
are clearly shifted towards lower integrated flux densities. 
}
{The newly detected methanol masers are mostly of low-luminosity and, except for some sources, 
their weakness is not due to distance effects or positional offsets. No specific correlation
is found with the physical parameters of the Hi-GAL clumps, except for sources with both CH$_3$OH and OH masers 
which tend to have higher mass and luminosity. The intensity of the methanol
masers correlates well with the velocity range of the maser emission, which suggests
that the low brightness of these masers is related to the number of maser spots in the
emitting region and their evolution with time.
}



\maketitle

\section{Introduction}
\label{sec:intro}

Our current understanding of the formation process of intermediate to massive
stars ($M \ga 5 M_\odot$) is limited, due to a combination of theoretical and
observational challenges.  On the theoretical side,
the formation of massive stars is a highly complex process, which is also
dependent on the interaction between the large-scale ($\ga 10\,$pc) structure of molecular clouds/clumps
and  dynamic fragmentation properties during the pre-stellar phase.

On the observational side, only very recently some progress has been made toward the
identification and study of the earliest phases of high-mass star formation (HMSF),
and the transition from the high-mass starless core (HMSC) phase (the likely
precursors of massive stars and clusters), to the high-mass proto-stellar
object phase. However, the physical conditions in the HMSC and the
exact evolutionary path from HMSC to massive stars are not  well constrained.

Maser lines from different molecular species,
including water, hydroxyl, and methanol, are
common observational phenomena associated with
massive star forming regions.
The relation between different types of masers found around
young stellar objects may yield important
information about the evolutionary
state of the regions (e.g., \citealp{szymczak2004}, \citealp{breen2010}).
In particular, since its discovery \citep{menten1991}, the 6.7-GHz methanol maser has been
recognized as one of the clearest signposts to the formation of young
high-mass stars. Theoretical models (e.g., \citealp{cragg2002})
and observational studies (e.g., \citealp{ellingsen2006}) suggest
that methanol masers are exclusively associated with  early phases of
massive star formation.  The methanol maser thus appears as an ideal tool
to detect a short-lived phase of HMSF, between the end of the large scale
accretion and the formation of massive proto-stars \citep{pestalozzi2012}.
However, it is difficult to establish the exact start of the methanol maser
phase and study the physical properties in that phase. 

A large sample of  massive cores can now be found in the {\it Herschel} infrared
GALactic Plane Survey (Hi-GAL), a key program  of the {\it Herschel}
Space Observatory to carry out a 5-band photometric imaging survey
at 70, 160, 250, 350, and $500\,\mu$m  of a $| b | \le 1^{\circ}$-wide strip of
the Milky Way Galactic plane \citep{molinari2010PASP}. This survey is now providing
us with large samples of dust clumps in a variety of evolutionary stages and in
various star-forming environments.

\section{
Observations}
\label{sec:obs}

\subsection{Previous observations}
\label{sec:prevobs}

During the last $10-15$ years, extensive 6.7-GHz methanol maser searches have been
undertaken using two different strategies: (1) targeted searches toward
colour-selected infrared  sources and known regions of star formation (e.g.,
\citealp{walsh1997}, \citealp{szymczak2002});
and, (2) unbiased surveys covering portions of the Galactic plane (for a summary
see \citealp{green2009}).  In particular, the Methanol Multibeam Survey
(MMBS, \citealp{green2009}) when completed will cover the whole Galactic plane in longitude,
and in a latitude range $|b| \le 2^{\circ}$. This survey also covers a previous
smaller survey carried out at Arecibo (the ``Arecibo Methanol Maser Galactic 
Plane Survey'', or AMGPS, \citealp{pandian2007a,pandian2011}). All of these surveys have
detected a total of about 800 masers in the range $\ell=20^{\circ} - 186^{\circ}$,
but many more ($\sim 1200 - 2500$) are expected to be found (\citealp{vanderWalt2005},
\citealp{pestalozzi2007}).

The targeted 1-$\sigma$ noise level of the MMBS is $\le 0.2\,$Jy \citep{green2008},
whereas the AMGPS yielded an rms noise level of $\sim 70\,$mJy \citep{pandian2007a}.
These are similar to the deepest previous unbiased surveys which had 1-$\sigma$ 
sensitivities between 0.09\,Jy and 1\,Jy (see \citealp{pestalozzi2005}). 
\cite{pandian2007b} found that the peak of the distribution of methanol masers 
as a function of flux density occurred between 0.9 and 3\,Jy. 
They also found a turnover in the number of sources at lower flux densities, 
but they could not  determine the shape of the distribution due to their completeness 
limit of 0.27\,Jy (the completeness limit in the MMBS is $\approx 0.8$\,Jy).  

Therefore, these surveys have been unable 
to detect a significant number of {\it low-flux density} (i.e., $\la 0.1\,$Jy or even $\ll 0.1\,$Jy) 
methanol masers, and thus we do not yet know how many such masers exist in the Galaxy 
and what is their physical nature.  In particular, in light of the association of
the 6.7-GHz methanol maser with the early stages of high-mass star formation 
(e.g., \citealp{pestalozzi2002}), it would be interesting to analyze 
if low-flux density masers can effectively be used to mark a specific 
evolutionary phase in high-mass star formation.

\subsection{Selection of the source sample}
\label{sec:sample}

The Hi-GAL survey offers the best opportunity to further study the issues described above,
since it allows us to look at large clump populations in various clouds with different
physical conditions, while using a self-consistent analysis to derive their physical
parameters (see, e.g., \citealp{elia2010, elia2013}, \citealp{olmi2013}).
Previous surveys suggest that methanol masers do not form towards low-mass
molecular clumps. Therefore, the mass of the Hi-GAL clump
can be used as the main selection parameter
to help identify new methanol masers.

For our observations at Arecibo, which were divided in three sessions
(July 2012, January 2013 and May 2013), we selected (and observed) a sample of 107 Hi-GAL
sources using the following basic criteria:
{\it (i)} the targets had to be located in the inner
Galaxy accessible to Arecibo (we limited the range to $\ell \sim 30^{\circ} - 60^{\circ}$);
{\it (ii)} the sources had a mass $M > 10 \,M_\odot$ (since the true distance to these
Hi-GAL sources had not been estimated yet, their preliminary masses
were calculated using an arbitrary distance of 1\,kpc);
{\it (iii)}  the sources must have been detected in
all Hi-GAL bands longward of $70\,\mu$m.
{\it (iv)} for the sources observed during the January and May 2013 sessions we also
checked that they were not near (within 1\,arcmin radius) any of the
already known methanol masers.
Note that given the observing restrictions at Arecibo, our selection criteria
intentionally used only mass as main parameter in order to have
at our disposal a large enough sample of sources.

\subsection{Arecibo observations}
\label{sec:arecibo}

The observations
were conducted with the 305$\,$m Arecibo Telescope\footnote{The Arecibo
Observatory is part of the National Astronomy and Ionosphere Center, which is
under a cooperative agreement with the National Science Foundation.}
in Puerto Rico, between July 2012 and May 2013 as we already mentioned.
We used the  C-Band High receiver to simultaneously observe the
$(5_1 - 6_0)$ transition of A$^+$ methanol at 6668.518-MHz
and the 6035.092-MHz ($^2\Pi_{3/2} J=5/2, F=3-3$) excited-state OH maser line.

We used the WAPP spectrometer, full Stokes polarization setup, 3-level sampling,
6.25$\,$MHz (280\,\kms) bandwidth, and 4096 channels per
polarization, resulting in a channel separation
of 1.53$\,$kHz (0.068\,\kms). We observed in ON-source (total power) mode, with
integration times of 5 minutes, which yielded a rms noise level
of $\simeq 5 - 10\,$mJy in each spectral channel, depending on the smoothing level.
Our sensitivity was thus much better than that achieved in the MMBS and AMGPS surveys.
The center bandpass LSR velocity was set to 70\,\kms.
The calibrator B2128+048 was observed in every run for
pointing and system checking (1$\,$min on-source observations).
%
%
The pointing was typically better than 10\sec. 
We measured a telescope beam size of
$\sim 42$\sec~(at 6.6$\,$GHz), and a typical
gain of $\sim6\,$K$\,$Jy$^{-1}$.

Data reduction was done in IDL\footnote{http://www.exelisvis.com/ProductsServices/IDL.aspx}
using specialized reduction routines
provided by the Arecibo Observatory. After checking
for consistency 
we subtracted low-order polynomial baselines.
The spectra were imported to CLASS\footnote{CLASS is part of the
GILDAS software package developed by IRAM.}
to measure line parameters and for further analysis.

%
%
%
\begin{sidewaystable*}
\vspace*{18cm}                
\caption{
6.7-GHz methanol masers detected with the Arecibo telescope.
$V_{\rm min}$ and $V_{\rm max}$ represent the minimum and maximum velocity corresponding to
the range of emission of the maser. $S_{\rm pk}$ represents the peak flux density, and
$\int S\, {\rm d}V$ is the integrated flux density in the velocity range $[V_{\rm min},V_{\rm max}]$.
$d$ is the estimated distance of the Hi-GAL source. The 8$^{\rm th}$ and 9$^{\rm th}$ columns show
the nearest Hi-GAL source (if the angular separation is $\le 200\,$arcsec) for crowded fields,
and the corresponding angular separation. In this case they are either the same source or there is
likely some contamination from the sidelobes.
The 11$^{\rm th}$ column indicates whether the maser
is a new detection (Y) or it is instead a known source (N). Source names in boldface indicate 
an OH maser counterpart determined using both positional and velocity association criteria (see Table~\ref{tab:listOH}).
}
\label{tab:listCH3OH}
\centering
\begin{tabular}{lccccccccccr}
\hline\hline
Name    &  RA  &  DEC  &  $V_{\rm min}$  & $V_{\rm max}$                              
        & $S_{\rm pk}$ & $\int S_{\it \nu}\, {\rm d}V$   & Nearest source  & Ang. separ.  & $d$  & New?  & Ref.\tablefootmark{a}  \\
        &  [J2000.0]     &  [J2000.0] &  [km\,s$^{-1}$] &  [km\,s$^{-1}$] & [Jy]   & [Jy\,km\,s$^{-1}$]  &   & [arcsec] & [kpc]   &  &  \\
\hline
%
G32.14+0.13  &  18:49:32.5  &  -00:38:09  &   92.3  &   93.2  &    0.03  &    0.01  &  G32.11+0.09 &           189 &    6.1 &  Y &  $-$  \\
G32.11+0.09  &  18:49:37.7  &  -00:41:01  &   90.2  &  104.5  &    1.16  &    0.77  &  G32.14+0.13 &           189 &    5.2\tablefootmark{b} &  N &  S1999  \\
{\bf G32.74-0.07}  &  18:51:21.8  &  -00:12:05  &   24.1  &   47.9  &   47.96  &  107.08  &  $-$ &  $-$ &    2.5 &  N &  CAS1995  \\
G33.09+0.06  &  18:51:30.5  &  00:10:41  &   77.9  &   84.9  &    0.14  &    0.09  &  $-$ &  $-$ &    5.3 &  Y &  $-$  \\
G32.82-0.08  &  18:51:32.1  &  -00:07:52  &   58.4  &   60.3  &    0.05  &    0.04  &  $-$ &  $-$ &    5.9 &  Y &  $-$  \\
{\bf G33.13-0.09}  &  18:52:07.9  &  00:08:14  &   70.4  &   82.2  &   11.36  &   14.47  &  $-$ &  $-$ &    4.9 &  N &  SHK2000  \\
G33.41-0.00  &  18:52:20.1  &  00:25:48  &   97.0  &  108.2  &    0.43  &    0.83  &  $-$ &  $-$ &  $-$ &  N &  SHK2000  \\
G33.59-0.03  &  18:52:46.0  &  00:34:10  &  102.6  &  103.5  &    0.02  &    0.01  &  G33.61-0.03 &           107 &  $-$ &  Y &  $-$  \\
G33.61-0.03  &  18:52:49.0  &  00:35:47  &  102.8  &  103.8  &    0.07  &    0.05  &  G33.59-0.03 &           107 &    6.5\tablefootmark{b} &  Y &  $-$  \\
G33.65-0.02  &  18:52:50.2  &  00:37:40  &  101.6  &  103.7  &    0.06  &    0.04  &  G33.61-0.03 &           114 &    4.5 &  Y &  $-$  \\
G34.37+0.23  &  18:53:13.6  &  01:23:31  &   54.9  &   63.7  &    1.63  &    0.99  &  $-$ &  $-$ &    1.6\tablefootmark{b} &  N &  SHK2000  \\
G34.08+0.01  &  18:53:30.5  &  01:02:04  &   54.7  &   61.6  &    0.73  &    0.55  &  $-$ &  $-$ &    3.7 &  N &  SKH2002  \\
G35.46+0.13  &  18:55:34.2  &  02:19:11  &   73.2  &   74.4  &    0.02  &    0.01  &  $-$ &  $-$ &    5.1 &  Y &  $-$  \\
G34.19-0.59  &  18:55:51.2  &  00:51:19  &   57.6  &   63.1  &    0.22  &    0.18  &  $-$ &  $-$ &    3.8 &  Y &  $-$  \\
G35.57-0.03  &  18:56:22.6  &  02:20:28  &  127.0  &  127.6  &    0.02  &    0.01  &  $-$ &  $-$ &   10.4 &  Y &  $-$  \\
{\bf G34.71-0.59}  &  18:56:48.2  &  01:18:46  &   77.8  &   80.0  &    0.01  &    0.00  &  $-$ &  $-$ &  $-$ &  Y &  $-$  \\
{\bf G35.13-0.74}  &  18:58:06.0  &  01:37:07  &   26.1  &   40.8  &   31.90  &   34.52  &  G35.14-0.75 & 62 &    2.2\tablefootmark{b} &  N &  SHK2000  \\
G35.14-0.75  &  18:58:09.9  &  01:37:27  &   26.2  &   39.4  &    1.70  &    1.80  &  G35.13-0.74 & 62 &    2.3 &  N &  SHK2000  \\
G36.42-0.16  &  18:58:23.2  &  03:02:11  &   71.4  &   72.3  &    0.03  &    0.01  &  $-$ &  $-$ &    8.6 &  Y &  $-$  \\
G36.83-0.02  &  18:58:39.0  &  03:28:01  &   52.7  &   64.5  &    2.51  &    6.35  &  $-$ &  $-$ &    3.9 &  N &  PGD2007  \\
{\bf G37.04-0.03}  &  18:59:04.2  &  03:38:34  &   77.9  &   86.5  &    9.56  &   17.61  &  $-$ &  $-$ &    5.6 &  N?\tablefootmark{c} &  SKH2002, PGD2007  \\
G37.34-0.06  &  18:59:43.1  &  03:53:40  &   51.3  &   52.6  &    0.02  &    0.01  &  $-$ &  $-$ &    9.8 &  Y &  $-$  \\
G37.19-0.41  &  19:00:43.4  &  03:36:24  &   29.4  &   30.1  &    0.07  &    0.02  &  $-$ &  $-$ &   11.1 &  Y &  $-$  \\
G37.86-0.60  &  19:02:36.0  &  04:07:04  &   49.3  &   54.2  &    0.19  &    0.25  &  $-$ &  $-$ &    3.4 &  Y &  $-$  \\
G38.93-0.36  &  19:03:42.0  &  05:10:24  &   31.0  &   33.8  &    0.04  &    0.05  &  $-$ &  $-$ &    2.7 &  N &  SHK2000  \\
G39.99-0.64  &  19:06:39.9  &  05:59:14  &   71.5  &   72.1  &    0.02  &    0.01  &  $-$ &  $-$ &    4.3 &  Y &  $-$  \\
G41.13-0.19  &  19:07:10.2  &  07:12:17  &   55.6  &   63.8  &    0.03  &    0.01  &  G41.16-0.18 &           106 &    4.3 &  Y &  $-$  \\
G41.16-0.18  &  19:07:11.2  &  07:14:02  &   55.6  &   63.6  &    0.07  &    0.08  &  G41.13-0.19 &           106 &    4.2 &  Y &  $-$  \\
G41.05-0.24  &  19:07:12.4  &  07:06:25  &   65.0  &   65.7  &    0.12  &    0.04  &  $-$ &  $-$ &    8.1 &  Y &  $-$  \\
G43.10+0.04  &  19:09:59.7  &  09:03:58  &    8.8  &   10.1  &    0.02  &    0.02  &  $-$ &  $-$ &   11.1\tablefootmark{b} &  Y &  $-$  \\
G43.53+0.01  &  19:10:52.9  &  09:25:44  &   51.6  &   52.9  &    0.09  &    0.03  &  $-$ &  $-$ &  $-$ &  Y &  $-$  \\
G47.04+0.25  &  19:16:41.5  &  12:39:20  &  101.5  &  102.0  &    0.02  &    0.00  &  $-$ &  $-$ &    4.7 &  Y &  $-$  \\
G45.87-0.37  &  19:16:42.9  &  11:19:10  &   59.6  &   60.5  &    0.02  &    0.01  &  $-$ &  $-$ &    5.2 &  Y &  $-$  \\
G46.32-0.25  &  19:17:09.0  &  11:46:24  &   41.5  &   41.9  &    0.02  &    0.01  &  $-$ &  $-$ &    7.4 &  Y &  $-$  \\
G56.96-0.23  &  19:38:16.8  &  21:08:07  &   29.3  &   30.6  &    1.12  &    0.42  &  $-$ &  $-$ &    3.0 &  Y &  $-$  \\
G59.78+0.63  &  19:41:03.0  &  24:01:15  &   36.2  &   40.6  &    0.03  &    0.03  &  $-$ &  $-$ &    2.1 &  Y &  $-$  \\
G59.63-0.19  &  19:43:49.9  &  23:28:37  &   21.9  &   32.8  &    0.58  &    0.51  &  $-$ &  $-$ &    2.3 &  Y &  $-$  \\
\hline
\end{tabular}
\tablefoot{
\tablefoottext{a}{S1999, \citet{slysh1999}. CAS1995, \citet{caswell1995b}. SHK2000, \citet{szymczak2000}. SKH2002, \citet{szymczak2002}. 
PGD2007, \citet{pandian2007a}.    }
\tablefoottext{b}{Distance determined from the BeSSeL Survey (see Section~\ref{sec:dist}).}
\tablefoottext{c}{This source has an angular distance of about 80\,arcsec from source 
G37.02-0.03 observed by SKH2002 and PGD2007, but the spectra are not similar. Emission partially overlaps only between 
78 and 80\,km\,s$^{-1}$. PGD2007 also detects some weak emission at about 85\,km\,s$^{-1}$.   }
}
%
\end{sidewaystable*}


%
%
\begin{table*}
\caption{
Same as Table~\ref{tab:listCH3OH} for the
6.0-GHz OH masers detected with the Arecibo telescope.
Source names in boldface indicate
a methanol maser counterpart (see Table~\ref{tab:listCH3OH}).
}
\label{tab:listOH}
\centering
\begin{tabular}{lcccccccr}
\hline\hline
Name    &  RA  &  DEC  &  $V_{\rm min}$  & $V_{\rm max}$                              
        & $S_{\rm pk}$ & $\int S_{\it \nu}\, {\rm d}V$  & $d$  & New?  \\
        &  [J2000.0]     &  [J2000.0] &  [km\,s$^{-1}$] &  [km\,s$^{-1}$] & [Jy]         & [Jy\,km\,s$^{-1}$]   & [kpc]  &  \\
\hline
%
{\bf G32.74-0.07}  &  18:51:21.8  &  -00:12:05 &   25.1  &   39.2  &    0.56  &   0.978  &    2.5 &  N  \\
G33.70+0.28  &  18:51:50.4        &  00:49:06  &   24.3  &   26.0  &    0.03  &   0.029  &    2.6 &  Y  \\
{\bf G33.13-0.09}  &  18:52:07.9  &  00:08:14  &   72.2  &   79.9  &    0.04  &   0.037  &    4.9 &  N  \\
G34.13+0.07  &  18:53:21.3        &  01:06:11  &   62.1  &   62.2  &    0.02  &   0.006  &    3.8 &  Y  \\
G35.74+0.15  &  18:56:01.0        &  02:34:34  &   81.8  &   85.8  &    0.02  &   0.031  &    5.6 &  Y  \\
G35.57-0.03  &  18:56:22.6        &  02:20:28  &   81.1  &   87.0  &    0.04  &   0.086  &   10.4 &  N  \\
{\bf G34.71-0.59}  &  18:56:48.2  &  01:18:46  &   81.8  &   85.0  &    0.02  &   0.031  &    $-$ &  Y  \\
{\bf G35.13-0.74}  &  18:58:06.0  &  01:37:07  &   32.8  &   37.1  &    3.92  &   3.615  &    2.2 &  Y  \\
G37.81+0.41  &  18:58:53.9        &  04:32:15  &   18.1  &   18.4  &    0.04  &   0.009  &    1.2 &  Y  \\
G35.29-0.89  &  18:58:57.0        &  01:41:40  &   57.6  &   58.5  &    0.04  &   0.019  &    2.5 &  Y  \\
{\bf G37.04-0.03}  &  18:59:04.2  &  03:38:34  &   81.1  &   85.1  &    0.05  &   0.073  &    5.6 &  Y  \\
G59.63-0.19  &  19:43:49.9        &  23:28:37  &   66.7  &   67.2  &    0.01  &   0.004  &    2.3 &  Y  \\
\hline
\end{tabular}
%
%
\end{table*}

%
\begin{figure}
\centering
\includegraphics[width=8.7cm,angle=0]{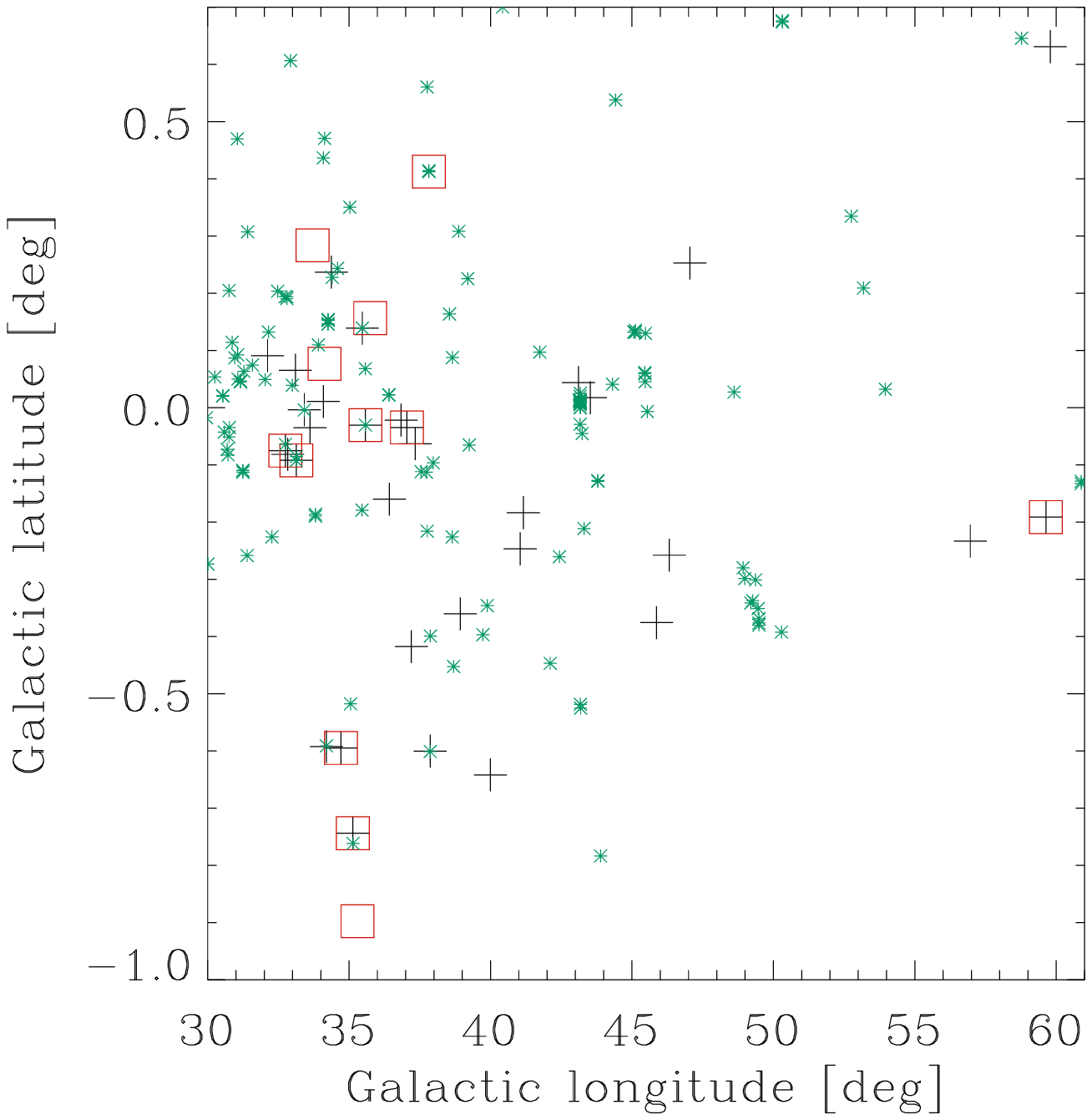}
\includegraphics[width=8.7cm,angle=0]{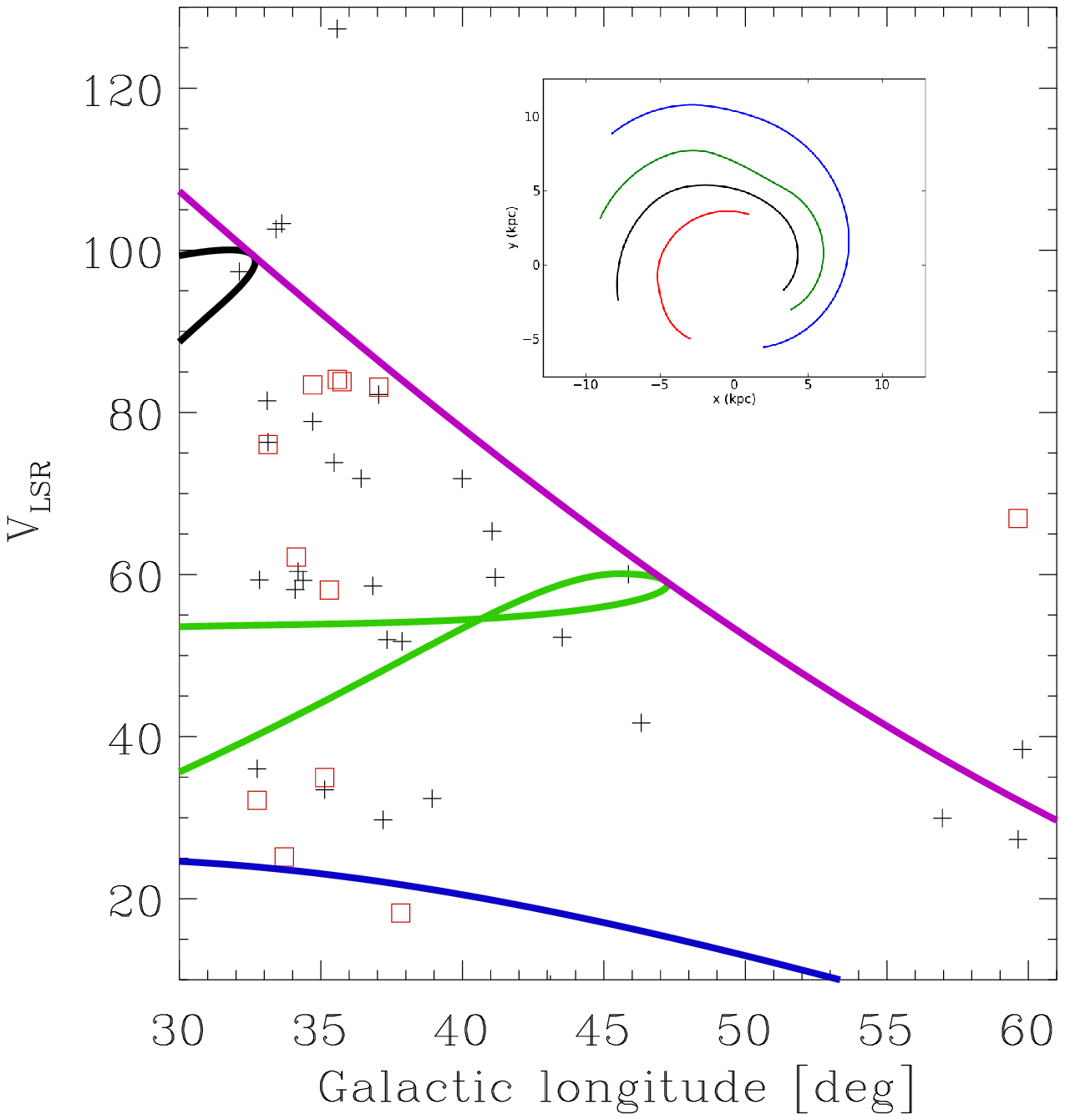}
\caption{
{\it Top panel}. Region searched for maser counterparts to Hi-GAL sources. Methanol
maser detections are denoted by "+" signs and OH maser detections are denoted by red squares.
Also shown are the UC \HII\ regions from the CORNISH catalog (green asterisks).
The concentration of sources around $\ell \simeq 35^{\circ}$ and $b \simeq 0^{\circ}$
is near the W44 region.
{\it Bottom panel}. Longitude-velocity distribution of all masers detected at Arecibo.
The coloured loci represent the spiral arms and
correspond to those shown in the inset: Norma (red), Scutum (black), Sagittarius (green),
Perseus (blue). The locus of the tangent point is shown in purple. 
}
\label{fig:maserpos}
\end{figure}

\section{Results}
\label{sec:results}

\subsection{Distance determination}
\label{sec:dist}

Assigning a distance to sources detected with a photometer is a crucial step in
the process of giving physical significance to all information
extracted from Hi-GAL data. While reliable distance estimates are
available for a limited  number of known objects (as, e.g., H{\sc II} regions, see \citealt{russeil2003},
and masers, see e.g. \citealt{green2011})
this information does not exist for the majority of Hi-GAL sources. We
adopted the scheme presented by \citet{russeil2011} aiming to assign
kinematic distances to large lists of sources: a $^{12}$CO (or
$^{13}$CO) spectrum (e.g., from the BU-FCRAO Galactic Ring Survey, 
or GRS, \citealp{jackson2006}) is 
extracted at the line of sight of every Hi-GAL source and the VLSR of the
brightest spectral component is assigned to it, allowing the
calculation of a kinematic distance (see details below). Using
extinction maps (derived from the {\it 2MASS} point source catalogue, see e.g. \citealp{schneider2011}) 
and a catalogue of sources with known distances (\HII\
regions, masers and others) the distance ambiguity is resolved and a
recommendation is given. In this way it is possible to produce a
``distance map'' having the same pixel size of the CO cube used to
extract the spectra for every target, where the value of the pixel is
the assigned distance of the Hi-GAL source(s) falling in that pixel. 

A source of error comes from the use of extinction maps to
solve for the distance ambiguity. A comparison between distances to
methanol maser sources assigned using extinction maps and HI
self-absorption, suggests that the former method tends to assign
more often the far heliocentric distance (Pestalozzi et al., {\it in prep.}).
The effects of this incorrect distance
assignment are more severe the larger the difference between near and
far heliocentric distances, because of the square dependence of mass
and luminosity on the source distance. For the present paper however, we have to rely on
the use of extinction maps for practical reasons and also because for
most of the sources no spectral line emission has been observed other
than what can be extracted from the GRS data cube \citep{jackson2006}.

We also checked if there was some overlapping between our sample of Hi-GAL clumps and the sources 
observed in the BeSSeL Survey\footnote{http://www3.mpifr-bonn.mpg.de/staff/abrunthaler/BeSSeL/} 
(``Bar and Spiral Structure Legacy Survey'', \citealp{brunthaler2011}).
We found only 4 methanol masers with a BeSSeL source within  5\,arcmin 
(an additional source is within 12\,arcmin), and in all these cases 
the distance determined by the BeSSeL team was {\it smaller} than that determined 
by us with the method described above, by a factor 0.19 to 0.85 (see \citealp{reid2009}). 
Clearly, if this trend will be confirmed, it will make 
these masers even weaker (see Section~\ref{sec:weak}).

\subsection{New methanol and OH masers}
\label{sec:new}

%
Out of a sample of 107 observed Hi-GAL sources
we detected a total of 32 methanol masers (30\% detection rate), with 22 sources being 
{\it new} detections, in the Galactic longitude range $[32^{\circ}.0,59^{\circ}.8]$.
We also detected 12 6.035-GHz OH maser (11\% detection rate), with 9 objects 
being new detections.  Only one of the newly detected methanol masers
(G34.71-0.59) has an associated OH maser.
The J2000.0 positions, velocity range of maser 
emission $[V_{\rm min},V_{\rm max}]$, peak flux density, $S_{\rm pk}$, 
integrated flux density, $\int S\, {\rm d}V$, and the estimated
distance, $d$, for each source are given in tables~\ref{tab:listCH3OH} 
and ~\ref{tab:listOH}, for the methanol and OH masers, respectively. 
Table~\ref{tab:listNODET} lists instead the Hi-GAL sources where no 
methanol maser was detected, and we report for each source the RMS of the
final spectrum.

The positions of the new detected methanol and OH masers are graphically shown in 
the top panel of Fig.~\ref{fig:maserpos}, where a higher concentration of sources is clearly seen 
around $\ell \simeq 35^{\circ}$ and $b \simeq 0^{\circ}$, where giant molecular clouds can be found that contain 
the W44 supernova remnant (e.g., \citealp{reach2005}). 
We note that a total of 7 maser sources are found outside the
galactic latitude range $|b| \leq 0^{\circ}.5$, which was explored by the 
AMGPS survey.  We also note that \cite{green2010} found that 97\% of 
their MMBS sources were at a latitude within $1^{\circ}$ of the Galactic plane.

As a comparison, Fig.~\ref{fig:maserpos} also shows the positions of the 
UC \HII\ regions, a typical signpost of  HMSF, from the CORNISH catalog 
(\citealp{hoare2012}, \citealp{purcell2013}). One can note that the methanol masers
follow the general distribution of the UC \HII\ regions, and in some cases their positions
are clearly coincident. In fact, we find that 7 methanol masers have a UC \HII\ region within 45\,arcsec
(i.e., about one Arecibo beam). Three of these masers (G32.74-0.07, G33.13-0.09 
and G33.41-0.00) were already known (see Table~\ref{tab:listCH3OH}), whereas the 
other four are new and lower flux density masers (G34.19-0.59, G35.46+0.13, G35.57-0.03
and G37.86-0.60). Although the nature of this association will have to be further investigated
through higher-angular resolution observations, we see that most of our newly detected methanol masers do
{\it not} have an associated UC \HII\ region (at the sensitivity level of the CORNISH catalog).
This result suggests that masers like these are more likely associated with the {\it pre}-UC \HII\ phase
of HMSF.

The bottom panel of Fig.~\ref{fig:maserpos} shows longitudes and velocities of 
all maser sources detected at Arecibo. 
Since 6.7-GHz methanol masers are only detected towards regions of HMSF 
(e.g., \citealp{pestalozzi2002}) they are expected to be found within spiral arms.
We find that the median (mean) velocity of all masers (methanol and OH) is
$60.0\pm20.7$\,km\,s$^{-1}$ ($61.9\pm20.8$\,km\,s$^{-1}$). Five sources
have velocities smaller than 30\,km\,s$^{-1}$, while there is only one source
with a velocity exceeding 120\,km\,s$^{-1}$.
By comparison with the velocity-longitude plot of \cite{pandian2007a}
we can see that we have a main group of sources, with $l\sim 35^{\circ}$ to
$40^{\circ}$ and $V_{\rm lsr} \sim 60$\,km\,s$^{-1}$ to 80\,km\,s$^{-1}$,
which fall near the Carina-Sagittarius arm. Another, less numerous group 
of sources, can be found near the overlapping region between the Carina-Sagittarius 
and Perseus arms. We also note that a significant fraction of the masers do 
not lie near any spiral arm loci, a phenomenon already discussed by
\cite{pandian2007b} and \cite{green2010}. This fact may be related with the
results of \citet{reid2009}, who found that on average the HMSF regions orbit the Galaxy 
$\approx 15\,$km\,s$^{-1}$ slower than the Galaxy spins.
%

Fig.~\ref{fig:spectraCH3OH1} shows that the spectra of the masers are composed of many
spectral features spread over a range of velocities. The total velocity
spread in an individual source depends on the sensitivity of the observation, particularly
when attempting to observe weak masers, and may also change as a result of intrinsic
variability of the components. The median (mean) spread in velocity for the methanol
and OH masers, respectively, is 2.8\,km\,s$^{-1}$ (5.4\,km\,s$^{-1}$)
and 3.1\,km\,s$^{-1}$ (5.0\,km\,s$^{-1}$). 
A large velocity range ($\ga 15-20\,$km\,s$^{-1}$) may also be caused by different
maser sources falling within the Arecibo beam. For example, the methanol maser source G32.82-0.08
has a component near 60\,km\,s$^{-1}$ and another near 30\,km\,s$^{-1}$, yielding
a velocity range of about 23\,km\,s$^{-1}$. However, the only new maser component is
the one at higher velocity, whereas the component at $\simeq 30\,$km\,s$^{-1}$ is likely
to be contamination from the known methanol maser G32.74-0.07, which has itself a large
velocity range.
Another example of contamination in the beam is described in the next section.


\subsection{Cross-scans and Long-integration observations}
\label{sec:long}

\begin{figure}
\hspace*{-4mm}
\includegraphics[width=10cm,angle=0]{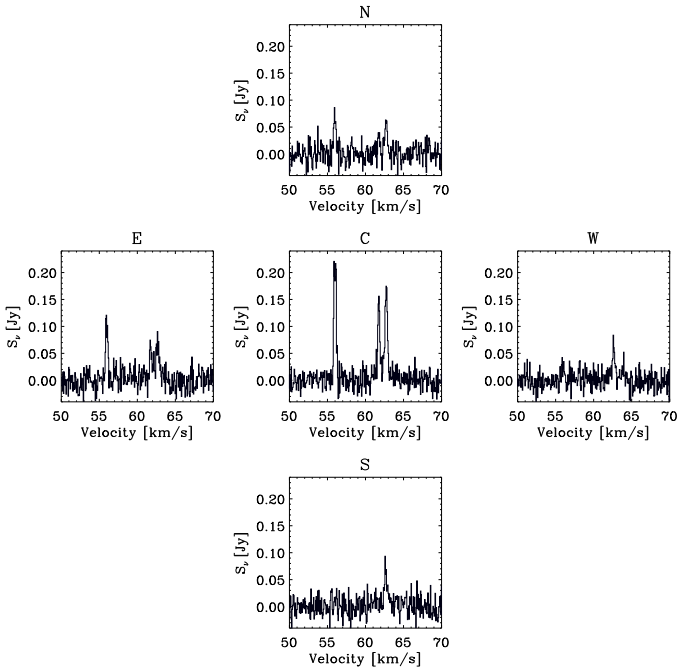}
\caption{
Cross-scan (with 24\,arcsec angular step) performed on source G41.16-0.18,
showing different velocity components and their variation as a function of
position.
}
\label{fig:cross}
\end{figure}

During our observations at Arecibo we
selected several sources (see Table~\ref{tab:cross})
to perform a cross-scan centered around the nominal Hi-GAL position
and with 24\,arcsec (about half beam) angular steps.
Table~\ref{tab:cross} shows the results obtained at various dates, and in
the case of source G41.16-0.18        
it also shows that different offsets were obtained depending
on the {\it velocity component} used for the computation (see Fig.~\ref{fig:cross}).
This is clearly an indication that distinct spatial maser components were simultaneously
present within the Arecibo beam, which can only be evidenced during a
cross-scan.
The results listed in Table~\ref{tab:cross} show that in most cases
the maser was observed within the main beam, and the estimated
offset was less than or comparable with
the Arecibo telescope pointing errors ($\la 15\,$arcsec, see Section~\ref{sec:arecibo}),
except on June $5^{\rm th}$, 2013, when the estimated offset was quite large.

An interesting question is that of multiple velocity maser components in the
low- and high-flux density masers (see Section~\ref{sec:clumps}). Therefore,
in a few selected low-flux density masers (G45.87-0.37, G43.53+0.01, G59.78+0.63)
we performed several consecutive 5\,min scans (totalling $15-25\,$min integration time),
that we then averaged in order to check for multiple maser components
that could have escaped the single 5\,min integrations because of
sensitivity limitations. The selected sources all initially appeared
to have a single component spectrum (G45.87-0.37, G43.53+0.01) or had
just a few components occupying a small ($\la 5\,$km\,s$^{-1}$) range
of velocities (G59.78+0.63). Our long-integration
spectra of these sources did {\it not} show  any new component.

%
%
\begin{table*}
\caption{
Results of cross-scans performed at Arecibo.
}
\label{tab:cross}
\centering
\begin{tabular}{lccccr}
\hline\hline%
Source    &  Date             & Velocity            & $\Delta$RA   & $\Delta$DEC        & Total    \\
          &  observed         & component           &              &                    & offset   \\
          &                   & [km\,s$^{-1}$]      &  [arcsec]    &  [arcsec]          & [arcsec] \\
\hline
G37.86-0.60  & 26-Jan-2013  & $51.1$              & $-0.7$       & 2.0                & 2.2  \\
G56.96-0.23  & 21-Jan-2013  & $29.8$              & 8.2          & 8.8                & 12.0 \\
G41.16-0.18  & 30-May-2013  & $56.0$              & 6.8          & 4.1                & 7.9 \\
G41.16-0.18  & 30-May-2013  & $62.8$              & 0.7          & $-2.7$             & 2.8 \\
G41.05-0.24  & 30-May-2013  & $65.4$              & 8.5          & $-11.1$            & 13.9 \\ 
G59.63-0.19  & 31-May-2013  & $29.6$              & 0.8          & $-6.0$             & 6.0  \\
G59.78+0.63    & 31-May-2013  & $38.3$              & 11.9         & 14.5               & 18.7 \\
G45.87-0.37  & 05-Jun-2013  & $59.9$              & 22.1         & 17.0               & 29.9 \\
%
\hline
\end{tabular}
\end{table*}

%

%
\begin{figure}
\hspace*{-7mm}
\includegraphics[width=9.7cm,angle=0]{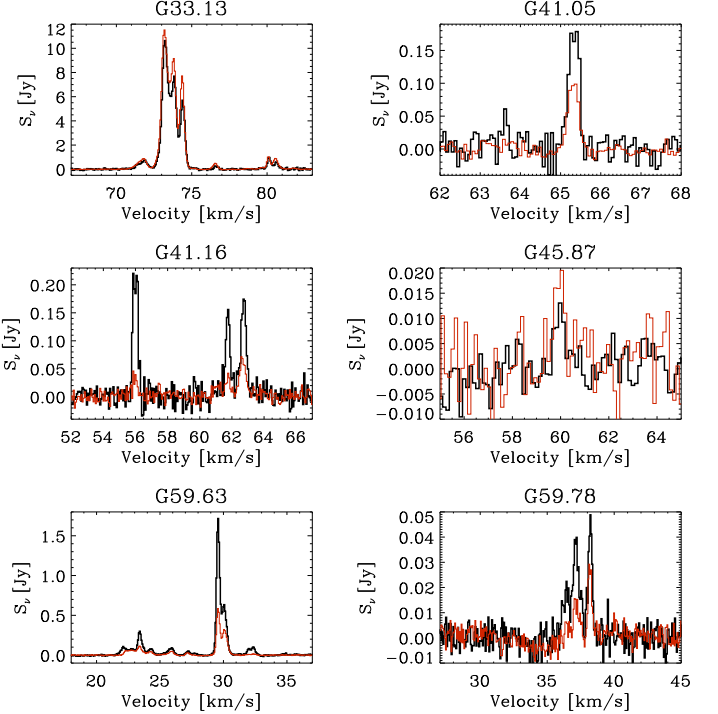}
\caption{
Spectra of methanol masers observed at different dates, with red (thin) and black (thick)
solid line representing older and more recent spectra, respectively.
}
\label{fig:var}
\end{figure}

In addition to sensitivity, another possible explanation for the non-detection
of multiple velocity components is that the source is being observed off-peak,
leaving only the most intense component detectable.
In this regard we note that sources G45.87-0.37 and G59.78+0.63 were also
observed in cross-scan mode, as described in the previous section and in
Table~\ref{tab:cross}. Therefore, the long-integration scans were performed
toward the observed peak position of the 5-pointings cross-scans. The observed
peak positions turned out to be very close to the estimated actual positions
of the sources. Despite this adjustment, we still did not observe any
additional velocity component.
This test toward a very small number of sources is neither complete nor conclusive,
and more sensitive observations toward a larger sample of low-flux density methanol masers
are clearly needed to resolve this issue.
However, we can use this result as a tentative indication that the weak
methanol masers detected at Arecibo do indeed tend to have fewer velocity
components than previously observed brighter masers (see also Section~\ref{sec:clumps}).

\subsection{Variability}
\label{sec:var}



In order to test the variability of the methanol masers detected by us, we have 
observed some of the sources at different dates, and we have selected both weak
and relatively bright masers.
In Fig.~\ref{fig:var} we show the methanol spectra observed in at least two
different dates (e.g., July 2012 or January 2013, and May or June 2013).
The sources showing the greatest variation in intensity are G41.16-0.16, G59.63-0.19,
G59.78+0.63 and, to a lesser degree, G41.05-0.24. The other sources show variations 
$\la 20\,$\% (see Table~\ref{tab:var}) which could be accounted for by calibration and 
pointing uncertainties.

In sources G33.13-0.09 and G41.16-0.16 we note the greatest difference between the
variation of the peak flux density and the total flux integrated over all velocity
components. Thus, not all of the observed maser components have varied
by the same amount during the period considered, either because they effectively
vary differently with time, or because they do not belong to the same source.
In the specific case of G41.16-0.16, given the results of the cross-scan performed
on this source (Section~\ref{sec:long}), we favour the second alternative.

%
%
\begin{table*}
\caption{
Variability of a few selected methanol masers. Columns n.5 and 6 list the percentage difference
of the peak flux density (at the velocity indicated in column n.4) and of the total flux between
the initial and final dates of observation.
}
\label{tab:var}
\centering
\begin{tabular}{lccccr}
\hline\hline%
Source         &  Initial Date     & Final date     & Velocity         & Flux density      & Flux               \\
               &  observed         & observed       & component        & variation         & variation          \\
               &                   &                & [km\,s$^{-1}$]   & [\%]              & [\%] \\
\hline
G33.13-0.09  & July 2012         & May 2013       & 73.3             & $-7.6$            & $-16.6$   \\
G41.05-0.24  & July 2012         & May 2013       & 65.4             & 41.6              & 34.9      \\
G41.16-0.18  & January 2013      & May 2013       & 56.0             & 379.5             & 134.6     \\
G43.10+0.04    & January 2013      & May 2013       & 9.5              & 9.1               & 8.3       \\
G45.87-0.37  & January 2013      & May 2013       & 60.0             & $-31.6$           & $-33.3$   \\
G59.63-0.19  & July 2012         & May 2013       & 29.6             & 202.0             & 158.8     \\
G59.78+0.63    & July 2012         & May 2013       & 38.3             & 96.3              & 107.4     \\
\hline
\end{tabular}
\end{table*}

%
\begin{figure}
\centering
\includegraphics[width=8.5cm,angle=0]{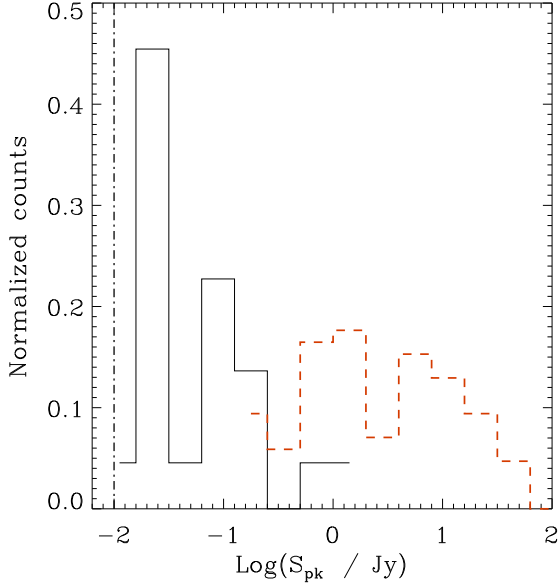}
\caption{
Histogram of the normalized counts (number of sources
per bin divided by the total number of sources) vs. peak flux for the new
methanol masers detected at Arecibo (black solid line) and for the AMGPS  (\citealp{pandian2007a,pandian2009}; red dashed line).
The rms noise level in our spectra was $\sim 5 - 10\,$mJy (shown by the dash-dotted vertical line)
in each spectral channel (see Table~\ref{tab:listNODET}).
}
\label{fig:histopkflux}
\end{figure}

%
\begin{figure}
\centering
\includegraphics[width=8.5cm,angle=0]{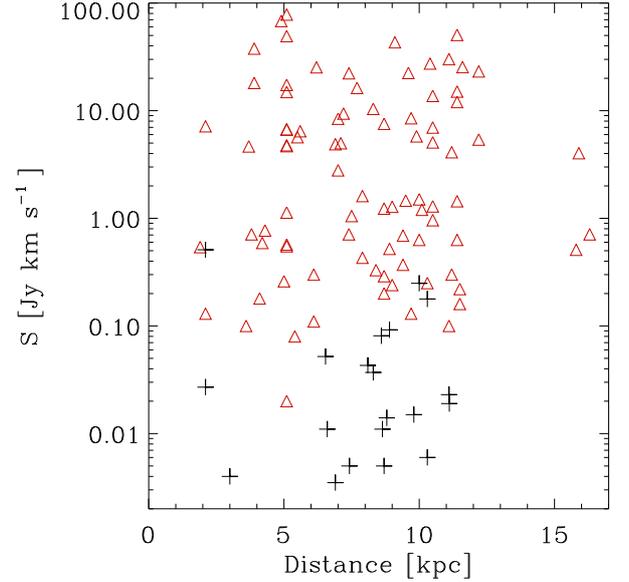}
\caption{
Plot of the methanol maser flux vs. distance of the associated clump for all
masers found by \cite{pandian2007a} (red triangles) and our {\it newly detected} methanol masers (black ``+'' signs).
The median value of the integrated flux density of
our own masers is $\simeq 0.03$Jy\,km\,s$^{-1}$, while the median value
for the  AMGPS masers is $\simeq 1.6$Jy\,km\,s$^{-1}$.
}
\label{fig:pandian}
\end{figure}


\section{Discussion}
\label{sec:discussion}

\subsection{Intrinsic maser intensity}
\label{sec:weak}

As we already mentioned in Section~\ref{sec:prevobs}, \cite{pandian2007b} 
found that the maximum of the distribution of their methanol masers
as a function of flux density occurred for peak flux densities between 0.9 and 3\,Jy.
Fig.~\ref{fig:histopkflux} shows that the peak flux density distribution 
of the 6.7-GHz methanol masers detected toward our sample of Hi-GAL sources  
does not follow the distribution found by \cite{pandian2007b}.
This is not surprising since our sensitivity is better than previous surveys,
and we have excluded from our analysis already known strong methanol masers.

But, clearly the interesting question is whether also the {\it intrinsic} intensity 
of these masers is lower than that of previously known methanol masers. In fact,
the simplest explanation of the weakness of our masers would be that, for example,
most sources in our Hi-GAL sample are systematically more distant than the sources 
observed by \cite{pandian2007a}.  However, we can exclude this observational
selection effect because Fig.~\ref{fig:pandian} clearly shows 
that although both source samples approximately cover the same distance range, 
the AMGPS masers are clearly shifted towards higher integrated flux densities, 
with a median value which is about 50 times higher than that of our sample. 
In addition, we have seen in Section~\ref{sec:dist} that where a BeSSeL counterpart
exists, its distance is smaller than that determined by us.

%
\begin{figure}
\centering
\includegraphics[width=8.5cm,angle=0]{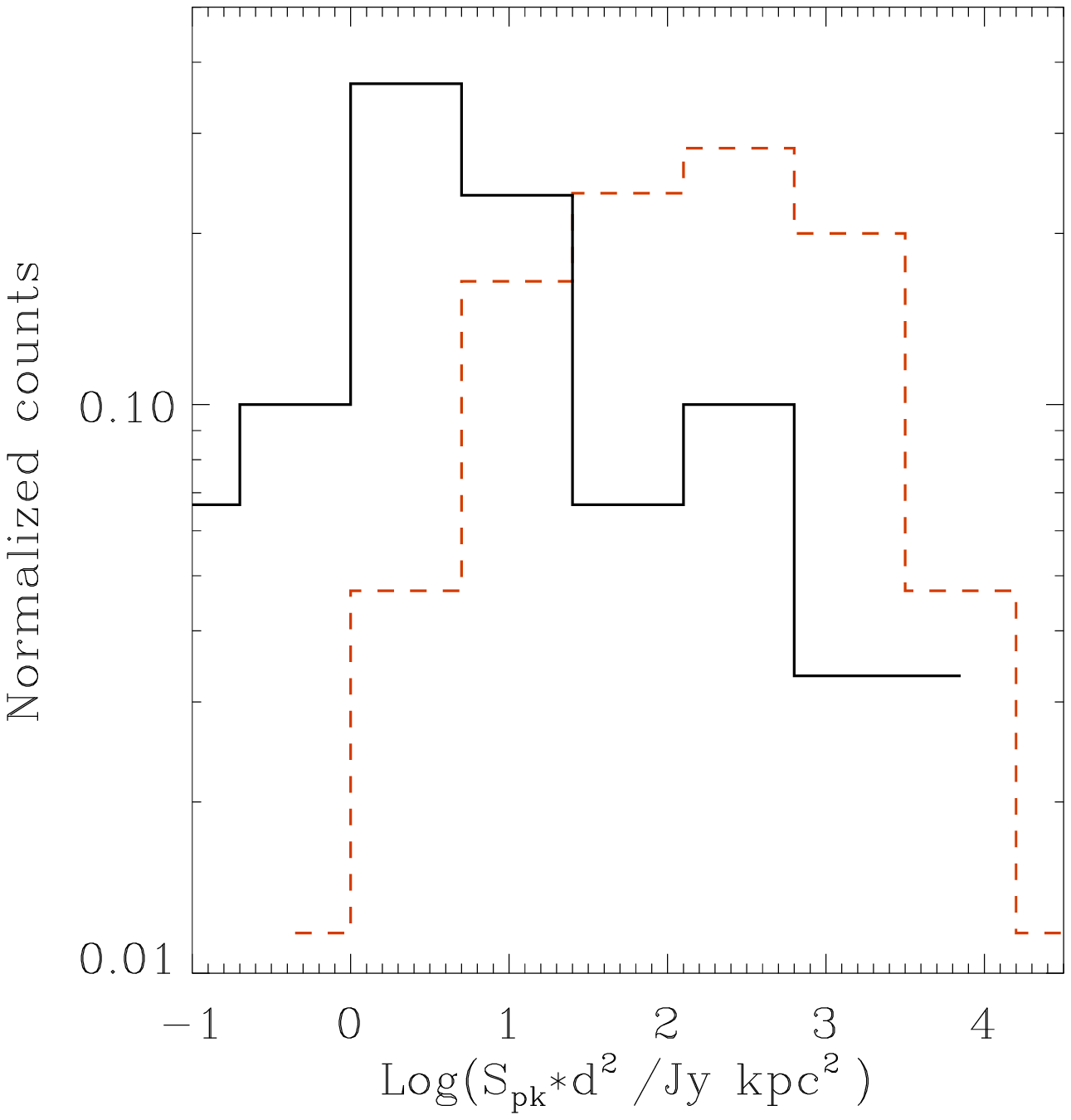}
\includegraphics[width=8.5cm,angle=0]{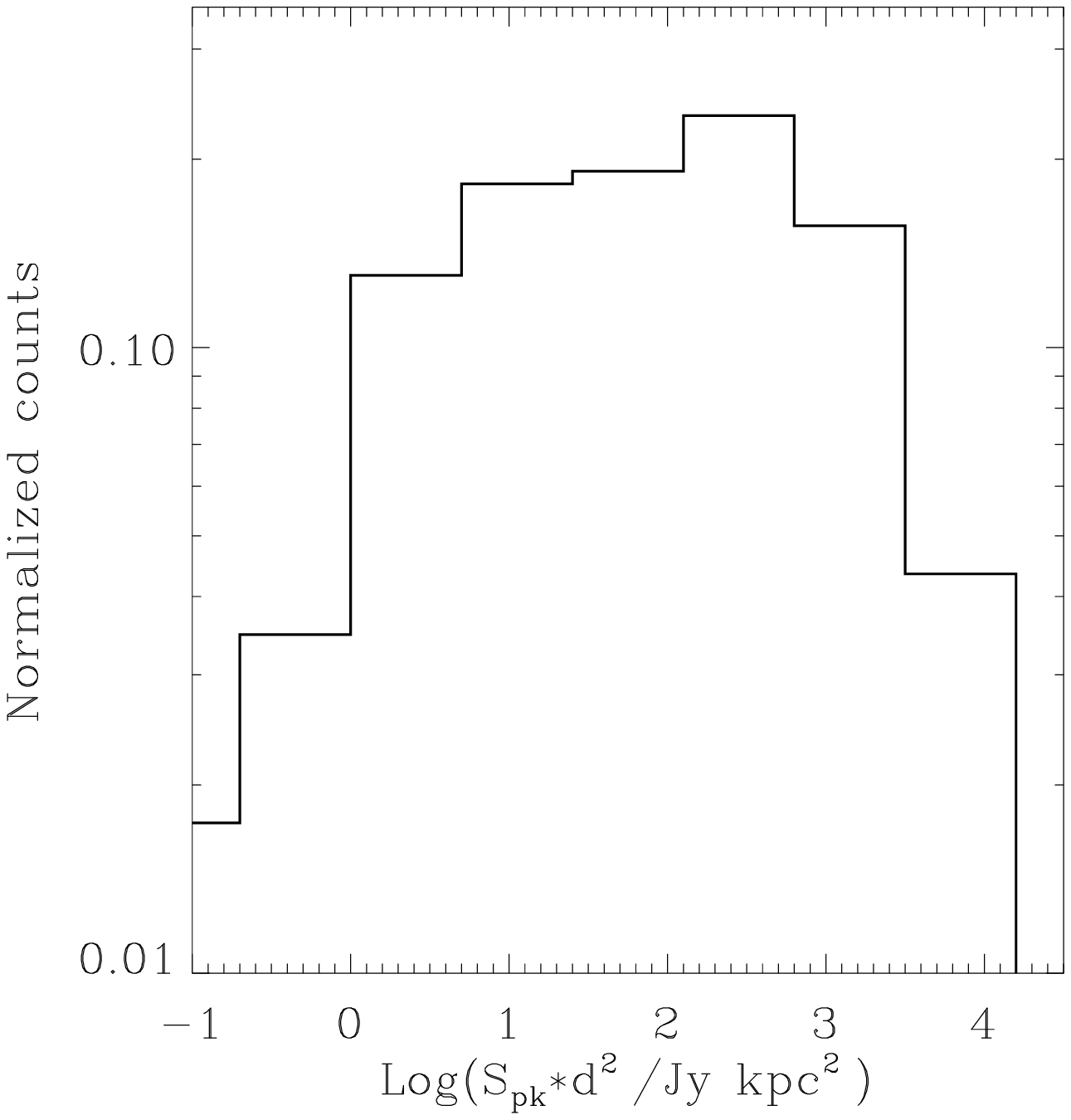}
\caption{
{\it Top panel.}
Histogram of the normalized counts (number of sources
per bin divided by the total number of sources) vs. $S_{\rm pk} \times d^2$,
for both the \cite{pandian2007a,pandian2009} data (red dashed line)
and our own results (black solid line), including new and known methanol masers.
{\it Bottom panel.} Histogram of the normalized counts vs. $S_{\rm pk} \times d^2$
of {\it all} methanol masers from both our survey and the AMGPS.
}
\label{fig:Fpkxd2}
\end{figure}

Another possibility to explain the difference shown in Fig.~\ref{fig:pandian},
would be to assume that the low brightness of our sources is caused by the masers
being systematically offset with respect to the nominal position of the
Hi-GAL source, or even by observing a known methanol maser in a sidelobe
of the Arecibo beam.  It seems very unlikely that {\it all} of the weak
masers detected by us have been observed with such a large pointing offset
to fully justify their lower intensity.
In fact, the cross-scans discussed in Section~\ref{sec:long} and listed
in Table~\ref{tab:cross}, show that the typical measured offset
may account for at most a $\simeq 25$\% decrease of the maser
peak intensity and thus cannot justify the difference of about a factor of
50 between the median values of the integrated flux densities previously mentioned.

%
\begin{figure}
\centering
\includegraphics[width=8.5cm,angle=0]{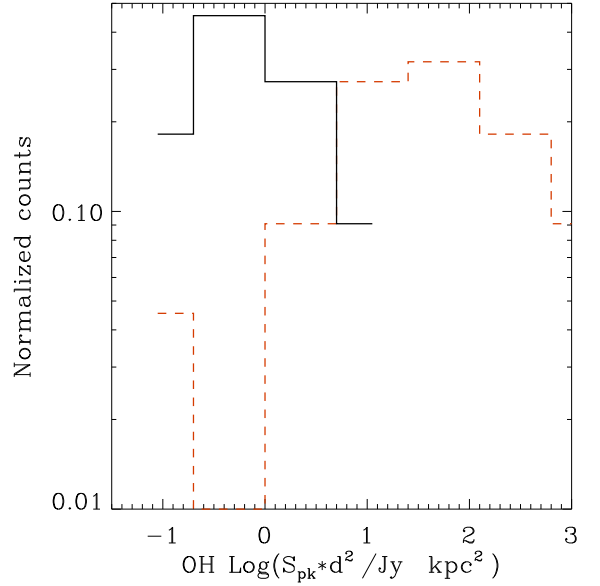}
\caption{
Histogram of the normalized counts (number of sources
per bin divided by the total number of sources) vs. $S_{\rm pk} \times d^2$
of the OH masers, for both the \cite{caswell1995a} data (red dashed line)
and our own results (black solid line). The bin width has been increased 
compared to Fig.~\ref{fig:Fpkxd2}.
}
\label{fig:Fpkxd2OH}
\end{figure}

Figure~\ref{fig:Fpkxd2} indeed suggests that
our methanol masers are intrinsically weaker than those detected in
in the AMGPS. In the top panel,
the normalized histogram of the peak flux density multiplied by
the distance squared, $S_{\rm pk} \times d^2$, shows a pronounced
peak at small values for our methanol masers. 
Given that the AMGPS and our samples have similar number of sources,
we do not expect statistical fluctuations to affect this comparison.
We also note that the lowest bins of the AMGPS may be affected
by completeness \citep{pandian2009}, 
whereas our distribution is robust even in the lowest bins,
since it is a pointed survey.

Figure~\ref{fig:Fpkxd2OH} represents a similar plot to that shown in
Fig.~\ref{fig:Fpkxd2}, but for the excited OH masers. In this case the 
comparison is done with the survey of \citet{caswell1995a}. 
In Fig.~\ref{fig:Fpkxd2} the distribution of the methanol masers detected 
by us has clearly its peak shifted toward smaller values of $S_{\rm pk} \times d^2$,
compared to the reference distribution. Figure~\ref{fig:Fpkxd2OH} also seems to
suggest a similar shift to lower intensities for our OH masers, but given the much 
lower number of detections this cannot yet be considered a firm conclusion.
Both figures~\ref{fig:Fpkxd2} and \ref{fig:Fpkxd2OH}, however, indicate that 
our blind survey toward Hi-GAL sources was indeed more sensitive to
the low-intensity tail of the distribution of methanol and OH maser intensities.


\subsection{Masers luminosity function}
\label{sec:lum}

\begin{figure}
\centering
\includegraphics[width=8.5cm,angle=0]{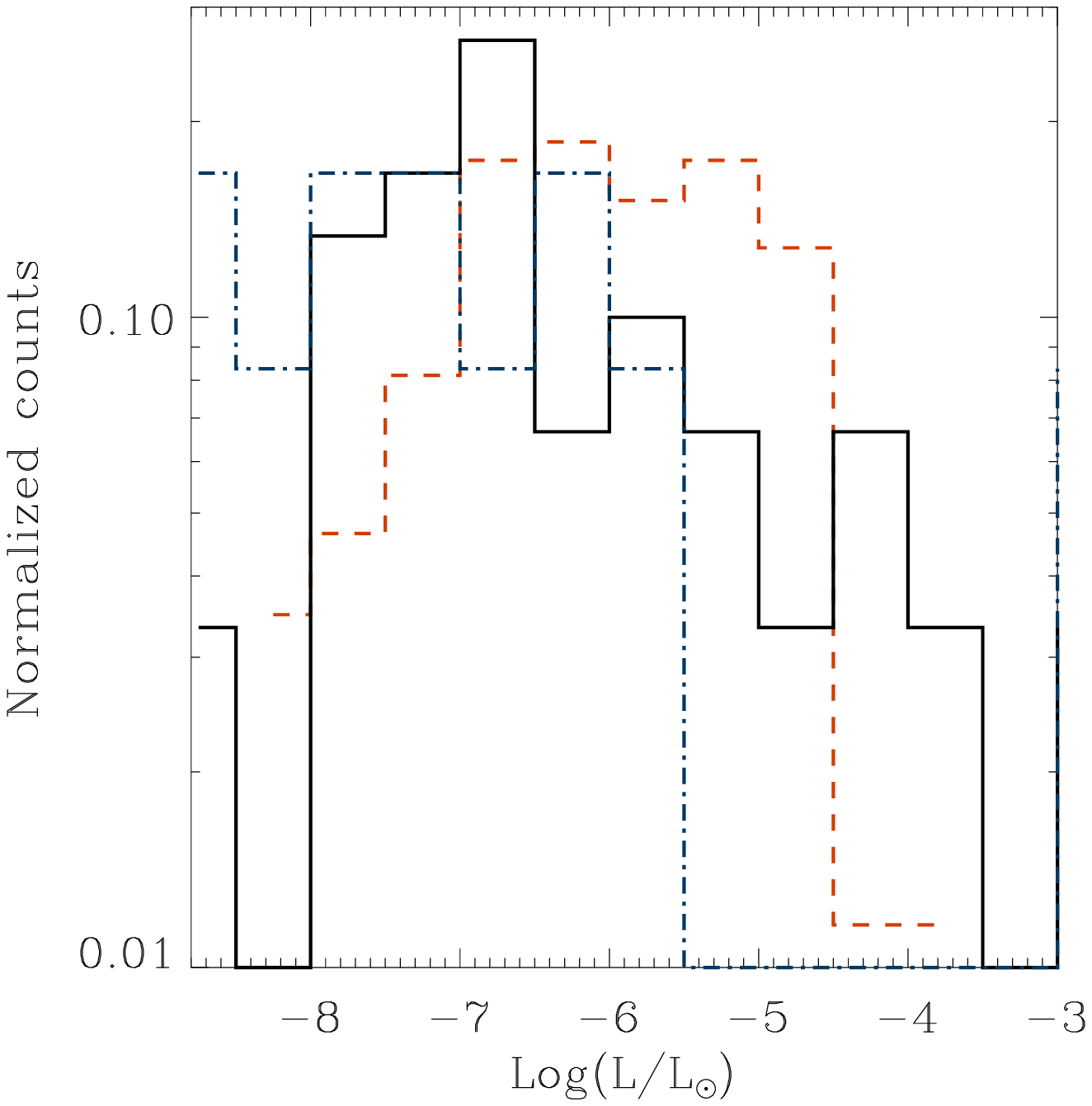}
\includegraphics[width=8.5cm,angle=0]{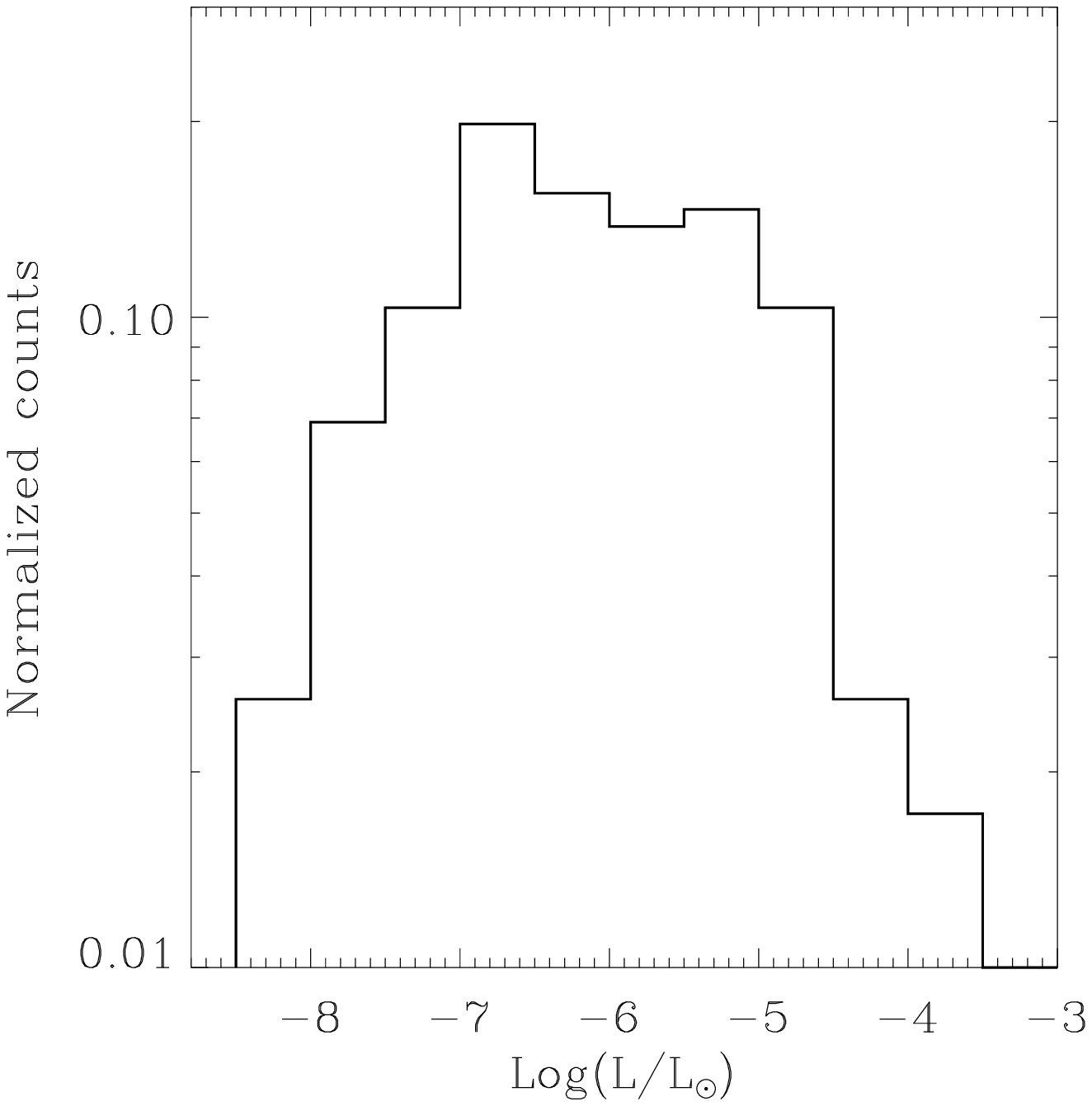}
\caption{
{\it Top panel.}
Histogram of the normalized counts (number of sources
per bin divided by the total number of sources) vs. luminosity of the methanol masers,
for both the \cite{pandian2007a,pandian2009} data (red dashed line)
and our own results (including new and known masers) with the luminosity of both
the methanol (black solid line) and OH masers (blue dash-dotted line).
{\it Bottom panel.} Histogram of the normalized counts vs. luminosity of {\it all}
methanol masers from both our survey and the AMGPS.
}
\label{fig:lum}
\end{figure}

Using the kinematic distance one can also calculate the isotropic 
luminosities of the masers (using their integrated flux densities, $S$) and 
the methanol maser luminosity function. The top panel of Fig.~\ref{fig:lum} shows the luminosity 
function of 6.7-GHz methanol masers, compared with the luminosity function of
the AMGPS masers \citep{pandian2009}. 
An interesting feature of Fig.~\ref{fig:lum} is that the peak of our distribution
nicely overlaps with the bins of the AMGPS distribution which may be affected
by completeness effects. Therefore, in the bottom panel of Fig.~\ref{fig:lum} 
we plot the distribution obtained by merging our sample with the AMPGS masers. 
We note that since the counts at luminosities $L < 10^{-7}\,L_\odot$
come mainly from our data, they are not affected by completeness effects and thus
the turnover at low luminosities (but still higher than the minimum measured luminosity) is real.
The shape of the merged distribution of the $S_{\rm pk} \times d^2$ parameter,
shown in the bottom panel of Fig.~\ref{fig:Fpkxd2}, is somewhat different
from the luminosity function, featuring an approximately flat-top with decreasing tails.

%

%
\begin{figure}
\centering
\includegraphics[width=7.0cm,angle=0]{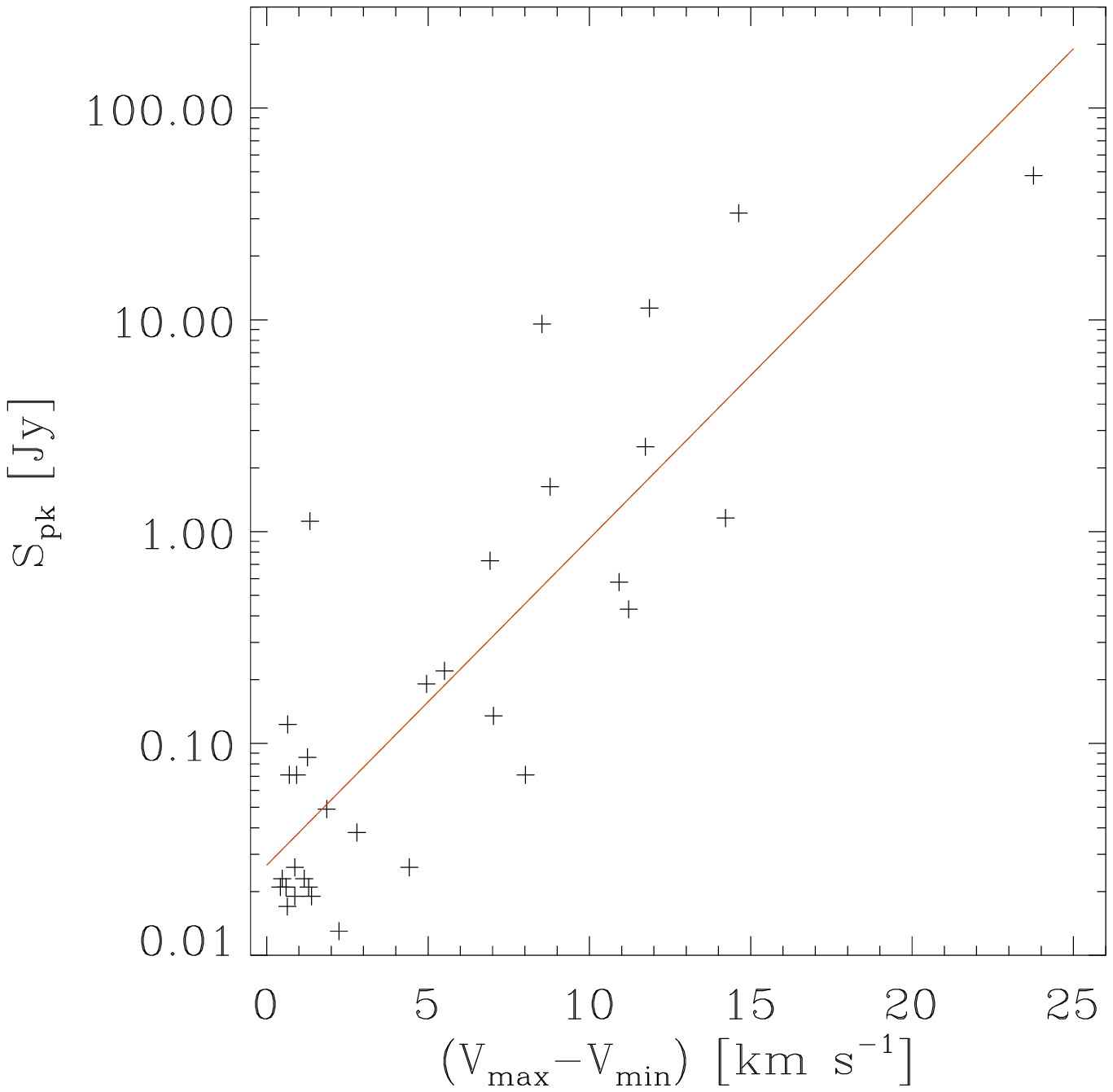}
\includegraphics[width=7.0cm,angle=0]{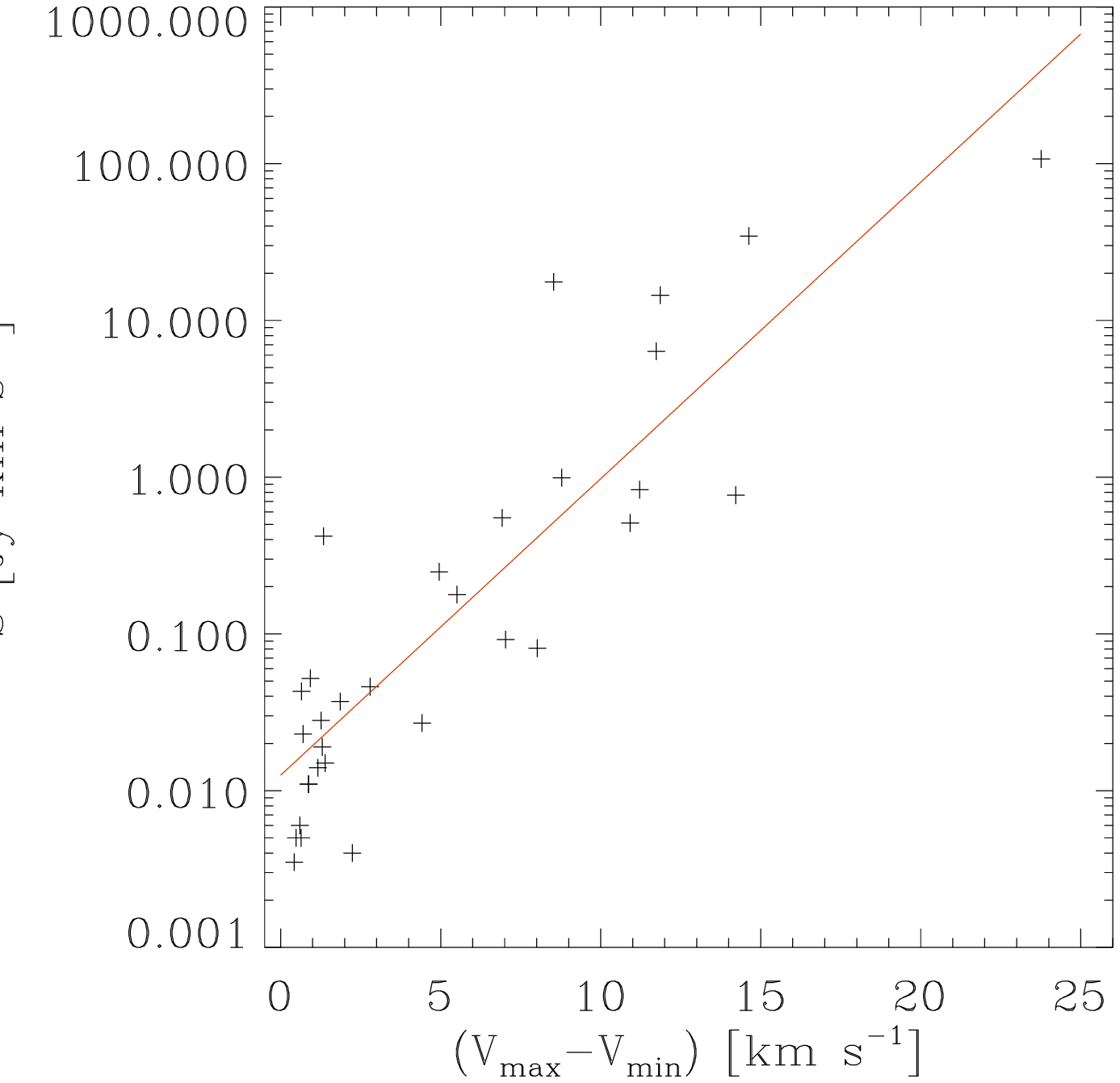}
\includegraphics[width=7.0cm,angle=0]{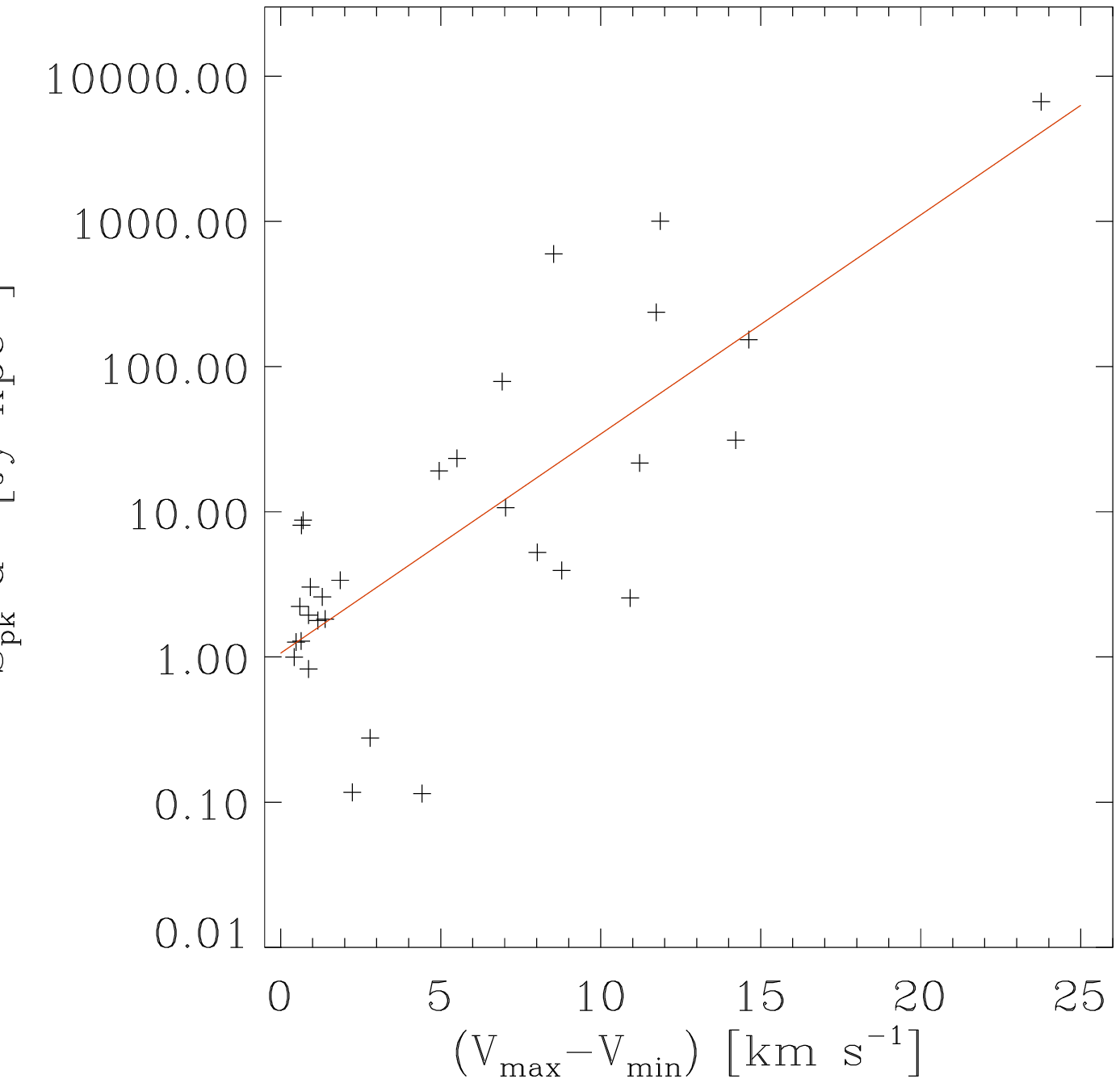}
\caption{
{\it Top panel.}
Plot of the peak flux density of the methanol masers vs. the difference
between the maximum and minimum value of the velocity range of emission.
The solid line represents the linear fit
(from Bayesian statistics, see text) to all points.
The Spearman rank coefficient is 0.75.
{\it Middle panel.} Same as above but with the integrated flux density, $S$,
plotted vs. the velocity range. The Spearman rank coefficient is 0.88.
{\it Bottom panel.} Same as above but with the $S_{\rm pk} \times d^2$ parameter
plotted vs. the velocity range. The Spearman rank coefficient is 0.69.
}
\label{fig:DV}
\end{figure}

While the luminosity functions and the distributions of the $S_{\rm pk} \times d^2$ 
parameter bear some resemblance, they are not expected to be exactly the same, since 
the latter does not take the linewidths and multiple emission components into account.
The luminosity of the maser 
emission is supposed to be a more reliable indicator of the physical conditions in a region 
as it depends on the conditions over a larger gas volume and will be less influenced by 
the fluctuations responsible for the intensity of a single spectral peak.
On the other hand, one might also think that the intrinsic sensitivity
of the quantity $S_{\rm pk} \times d^2$ to the main (or single) component of the 
maser emission would make this parameter better suited for a comparison of the relative strength 
between our maser sample and the AMGPS masers. 

For example, this might be the case if
the luminosity of the weak masers ($S_{\rm pk} \ll 1\,$Jy) were systematically underestimated
because the sensitivity of the observations is not good enough to detect all multiple emission
components. The top panel of Fig.~\ref{fig:DV} would appear to support this assumption, 
since the masers with the lowest $S_{\rm pk}$ are also characterized by a lower velocity range, 
or actually show only one emission component.
However, the middle panel of Fig.~\ref{fig:DV} shows that also the integrated flux density
follows the same behaviour. In both cases we have used the Bayesian IDL routine {\sc LINMIX\_ERR} 
to perform a linear regression, to find the slope of the best fit line. 
The relatively high correlation for the plots in the top and middle panels of Fig.~\ref{fig:DV} 
suggests that the integrated flux density is mostly determined by the most intense emission component.

Therefore, the luminosity function, or the distribution of the $S_{\rm pk} \times d^2$ parameter,
of the methanol masers found toward our sample of Hi-GAL dust clumps, shows a markedly different 
shape compared to that, for example, of the
AMGPS masers. The luminosity function is also different from that estimated by \cite{pestalozzi2007} 
for the whole Galaxy. Although our sample is admittedly still small, this result
is suggesting that the luminosity of the methanol masers detected towards our sample
of Hi-GAL sources is not distributed according to the luminosity function observed in 
unbiased surveys.
Fig.~\ref{fig:lum} also shows the luminosity function of the 6.0-GHz OH masers, which is 
more uncertain due to the low number of OH masers detected. The range of OH luminosities 
is similar to that of the methanol masers, but no other trend is visible,
and a more detailed comparison with the luminosity function of the methanol 
masers cannot be done with these data alone.

\subsection{Properties of associated Hi-GAL clumps}
\label{sec:clumps}

Previous works have already attempted to find possible correlations between the physical
parameters of the gas/dust clump where maser activity is present and, e.g., the 
luminosity of the maser emission.  \cite{breen2010}, for example, found that the
1.2-mm dust clumps with associated methanol masers have higher values of 
mass and radius than those with no associated 6.7-GHz methanol maser. 

%
\begin{table*}
\caption{
Median values of peak and integrated flux density for {\it all} detected methanol masers.
}
\label{tab:summary}
\centering\begin{tabular}{lccccccr}
\hline\hline
%
             \multicolumn{2}{c}{All detections}                   &    & \multicolumn{2}{c}{Sources with OH maser }    &
& \multicolumn{2}{c}{Sources without OH maser }  \\
\cline{1-2}
\cline{4-5}
\cline{7-8}
Median $S_{\rm pk}$  & Median $\int S\, {\rm d}V$  &  & Median $S_{\rm pk}$
& Median $\int S\, {\rm d}V$  &  & Median $S_{\rm pk}$   & Median $\int S\, {\rm d}V$  \\
$[{\rm Jy}]$  & [Jy\,km\,s$^{-1}$] &  & [Jy] & [Jy\,km\,s$^{-1}$]  &  & [Jy] & [Jy\,km\,s$^{-1}$] \\
\hline
%
        0.09     & 0.05             &   & 11.4     & 17.6         &    & 0.07   & 0.04  \\
%
\hline
\end{tabular}
\end{table*}

%
\begin{figure}
\hspace*{-4mm}
\includegraphics[width=9.7cm,angle=0]{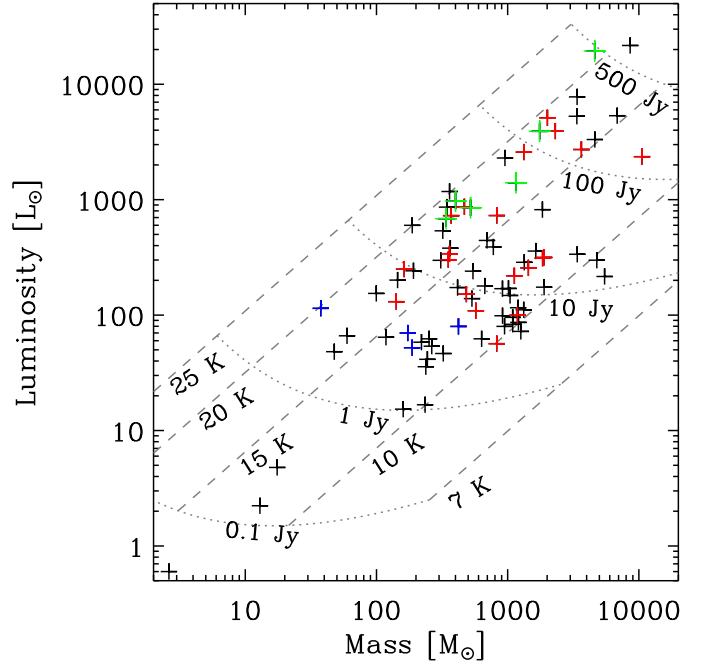}
\caption{
Luminosity versus mass for all clumps (black ``+'' signs) observed at Arecibo,
estimated using the distances listed in Table~\ref{tab:listCH3OH}.
Sources where a methanol or OH maser has been detected are shown in red and blue,
respectively; sources with both masers are shown in green (both newly detected
and previously known masers are shown).
The dashed lines are loci at constant T = 10, 20, 30, and 40 K.
Roughly orthogonal to these are loci (dotted lines) of constant 250$\,\mu$m
flux density, ranging from 0.1 to 500\,Jy (assuming a modified blackbody spectral energy distribution 
with $\beta = 1.5$ and a fixed, ``typical'' distance of $\sim 6\,$kpc).
}
\label{fig:LMmasers}
\end{figure}

In Fig.~\ref{fig:LMmasers} we present a luminosity vs. mass plot of all the Hi-GAL
sources observed by us at Arecibo for which luminosity and mass could be calculated. 
These two physical parameters were determined using the estimated distance to the
sources (see Section~\ref{sec:dist}) and using a simple, single-temperature 
spectral energy distribution (SED) model to fit the PACS and SPIRE flux densities.
Therefore, both luminosity and mass refer to the cold ($T_{\rm d} \la 20-30\,$K) 
dust envelope of the Hi-GAL clumps (and should therefore be indicated as $M_{\rm env}$ and $L_{\rm env}$),
and may be insensitive to warmer dust (emitting mostly shortward of $\sim 24\,\mu$m)
if a central protostar (or cluster of protostars) has already formed within the clump.

In Fig.~\ref{fig:LMmasers} one should thus be aware that the envelope luminosities may be
{\it underestimating} the total (i.e., bolometric) clump luminosity, $L_{\rm bol}$, 
if both warm and cold dust components exist within the Hi-GAL clump.  With this {\it caveat}, 
in this figure we then show in different colors the
groups of sources associated with a methanol or OH (or both) maser.
It can be noted that while sources with only a  methanol maser associated
do not show any recognizable distribution pattern, sources with either just
an OH maser or with both masers seem to occupy different regions in this plot.
Hi-GAL clumps where {\it both} masers have been detected do show somewhat
higher masses and luminosities.
However, given our low number of sources and the fact that the internal structure of the Hi-GAL clumps 
is not yet known, these trends are not statistically significant.

\begin{figure}
\centering
\includegraphics[width=9.0cm,angle=0]{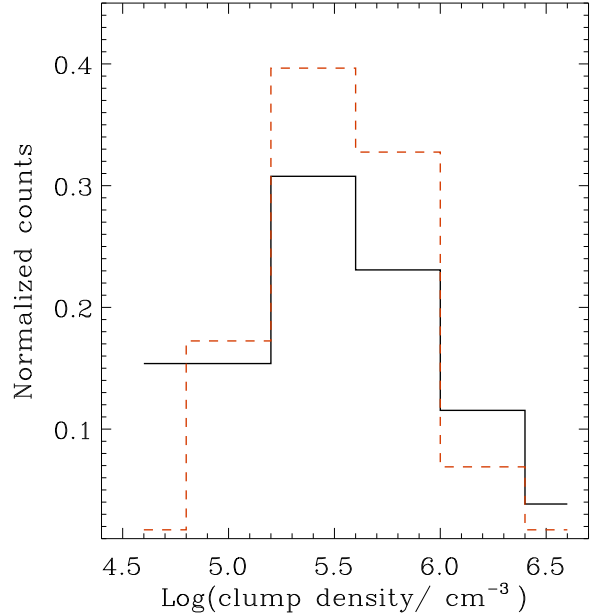}
\caption{
Histogram of the density of {\it all} observed Hi-GAL clumps in our survey.
The black (solid) and red (dashed) lines represent the distribution of sources with
and without, respectively, an associated methanol maser.
}
\label{fig:density}
\end{figure}

Likewise, for these weak masers
we are unable to find any evidence of the trends discussed by \cite{breen2010}.
We note that their analysis is based on the data by \cite{hill2005},
whose angular resolution is comparable to that of the Hi-GAL maps.
We have also estimated the gas density in the observed Hi-GAL clumps,
with and without an associated methanol maser. The results are plotted
in Fig.~\ref{fig:density}, which shows that no significant difference is
observed between the distributions of sources with and without a maser.
These results are not too surprising, since dust clumps at distances $\ga 1\,$kpc,
observed at relatively low angular resolution, may be still large enough
to host more than one compact source, possibly in different evolutionary phases.
Therefore, an improved analysis of the correlation between maser activity
and the physical properties of the gas clump will be possible only
through higher angular resolution maps of the molecular gas
in the masers environment.

\subsection{OH association}
\label{sec:oh}


As we mentioned in Section~\ref{sec:new} only one of the new methanol
masers discovered by us has an associated excited OH maser, when the velocity range of the maser emission
is also used as a criterion for the masers to be physically associated.
However, if we also include the known methanol masers, then the number of sources with
both maser types is 5 out of a total of 32 methanol masers.
The main observational property that characterizes
the sources where both types of maser activity is present, is the higher intensity of
the methanol masers as compared to sources with no OH detection. In fact,
Table~\ref{tab:summary} is tentatively suggesting (because of the large scatter) that
both peak flux density and flux of methanol maser emission may have on-average
higher values in sources associated with an OH maser.
Although our sample is relatively small, and the scatter
around the average values too high, this result is consistent with the findings
of \cite{breen2010} toward a larger sample of more intense methanol and OH
masers. We should note that
the situation is not similar when one considers the OH masers (see Table~\ref{tab:summaryOH}).
In fact, the median values of the peak flux density of the OH masers is comparable in sources with
and without an associated methanol maser.
Despite the {\it caveats} mentioned above, we can discuss our results
in the context of maser excitation and compare them with proposed scenarios of
maser evolution.

The model calculations of \cite{cragg2002}
showed that the coincidence of OH and methanol masers in many sources can be explained
in terms of common excitation conditions which produce population inversions
simultaneously in both molecules. The masers require infrared pumping radiation
from warm ($T_{\rm d} > 100\,$K) dust and are most likely to form in cooler
($T_{\rm k} < 100\,$K) gas of moderately high density ($ 10^5 < n_{\rm H} < 10^{8.3}\,$cm$^{-3}$).
When methanol and OH masers are detected, it is necessary that both be present at high
abundance in the gas phase. When masers of one or other molecule are seen in
isolation, \cite{cragg2002} gives two possible explanations; either the non-masing
molecule is not sufficiently abundant, or the local conditions produce maser action
in the favoured molecule alone.

These authors show that the 6.7-GHz methanol maser
is excited at relatively lower densities ($n_{\rm H} > 10^4\,$cm$^{-3}$), compared to
the 6.035-GHz OH line (which requires $n_{\rm H} \ga 10^5-10^6\,$cm$^{-3}$), and is independent of gas
density up to $n_{\rm H} \sim 10^8\,$cm$^{-3}$. This indicates that the methanol maser
excitation mechanism is predominantly radiative. On the other hand, the excitation of OH
is more sensitive to the local density and also extends to higher densities, where the
methanol maser is instead subject to collisional quenching. \cite{cragg2005}, however,
showed that collisional quenching at high density of the 6.7-GHz methanol maser
becomes less probable when new rate coefficients are used.
Furthermore, it should be noted that
the model calculations of \cite{cragg2002} all refer to emerging masers with a brigthness
temperature exceeding $10^4\,$K, or 0.1\,Jy for a 6-GHz maser of size 0.7\,arcsec. Therefore,
their conclusions may not be entirely valid for weaker maser emission.

%
\begin{table*}
\caption{
Median values of peak and integrated flux density for {\it all} detected OH masers.
}
\label{tab:summaryOH}
\centering\begin{tabular}{lccccccr}
\hline\hline
%
             \multicolumn{2}{c}{All detections}                 &  & \multicolumn{2}{c}{Sources with CH$_3$OH maser }    &     &
\multicolumn{2}{c}{Sources without CH$_3$OH maser }  \\
\cline{1-2}
\cline{4-5}
\cline{7-8}
Median $S_{\rm pk}$  & Median $\int S\, {\rm d}V$  &  & Median $S_{\rm pk}$
& Median $\int S\, {\rm d}V$  &  & Median $S_{\rm pk}$   & Median $\int S\, {\rm d}V$  \\
$[{\rm Jy}]$   & [Jy\,km\,s$^{-1}$] &  & [Jy] & [Jy\,km\,s$^{-1}$]  &  & [Jy] & [Jy\,km\,s$^{-1}$]          \\
\hline
%
        0.04    & 0.03              &   & 0.05     & 0.07             &    & 0.03   & 0.02  \\
\hline
\end{tabular}
\end{table*}

%

According to \cite{cragg2002, cragg2005}, there is therefore a limited range of conditions 
which favour maser action in just one molecule, and they claim that molecular abundance is 
likely the determinant factor of methanol and OH maser activity in HMSF regions. If that is
indeed the case, then the presence or absence of both maser types should be able to be tied
with the chemical evolution and age of the HMSF region. 
A common proposed scenario (\citealp{cragg2002}, \citealp{ellingsen2011}  and references therein) 
is that the gas-phase methanol abundance is enriched in maser regions following the
evaporation of icy grain mantles. The same process is responsible for injecting water molecules
in the gas-phase, then leading to production of OH through photodissociation or ion-molecule
chemistry. This is consistent with a time-span when both masers are present, but chemical models
(e.g., \citealp{charnley1992, charnley1995}) predict that the OH abundance 
should peak after the methanol
abundance has started to diminish (about $10^5\,$yr after grain mantle molecules have been
injected in the gas-phase). Therefore, methanol masers are expected to start (and finish) 
earlier than OH masers.
We finally note that 2 of the 5 sources with double maser emission also show a MIPS 24$\mu$m
counterpart. 

\subsection{IR counterparts}
\label{sec:ir}

The association between mid-infrared emission and 6.7-GHz methanol masers has already been
investigated in several of previous surveys. Recent works (e.g.,  \citealp{ellingsen2006}, 
\citealp{cyganowski2009}, \citealp{pandian2011}) have found a very close correspondence between 
6.7-GHz methanol masers and mid-infrared emission, although it should be noted that in
some cases (\citealp{ellingsen2006}, \citealp{cyganowski2009})  these were targeted searches 
toward GLIMPSE point sources that resulted in high detection rates of 6.7-GHz methanol masers.
However, even \cite{pandian2011} found that almost all AMGPS 6.7-GHz methanol masers 
had indeed associated  a MIPS 24$\mu$m counterpart within 5\,arcsec.

\begin{figure}
\centering
\includegraphics[width=9.0cm,angle=0]{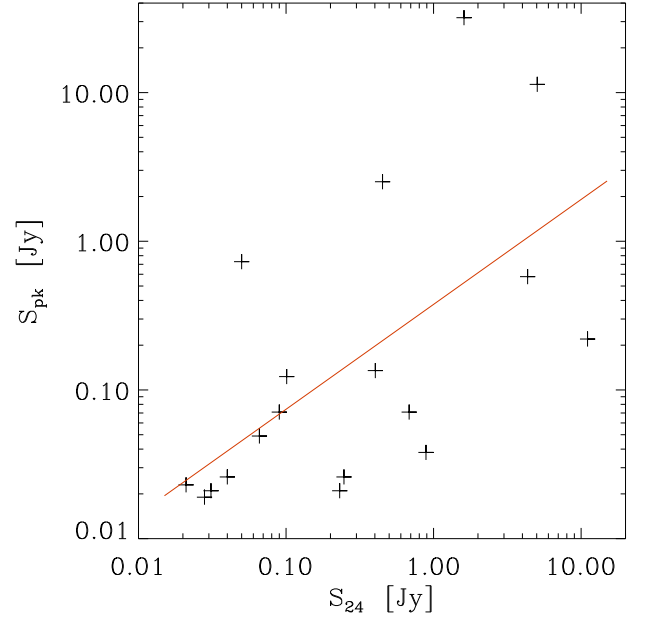}
\caption{
Plot of the peak flux density of all methanol masers associated with a MIPS 24$\mu$m counterpart
vs. the flux density of the MIPS source. The solid line represents the linear fit
to all points.  The Spearman rank coefficient is 0.57.
}
\label{fig:mips}
\end{figure}

We have thus searched for MIPS 24$\mu$m counterparts associated with our Hi-GAL sources,
and found that 18 out of 32 methanol masers have a mid-infrared source within 5\,arcsec
(the median angular separation is 2.5\,arcsec). Of the 22 new methanol masers 13 also 
have an associated MIPS 24$\mu$m counterpart, hence
in our sample the fraction of sources with (IR-loud) or without (IR-quiet)  
a MIPS 24$\mu$m counterpart is about the same. 
We find no statistically significant difference between, for example,
the median values of the $S_{\rm pk} \times d^2$ parameter associated with either
the IR-loud or IR-quiet sub-groups. However, Fig.~\ref{fig:mips} shows that, despite
the large scatter, there seem to be a moderate correlation between the peak flux density of the 
methanol masers and the flux density of the associated MIPS 24$\mu$m counterpart,
which will need to be confirmed by future (high-angular resolution) observations. 
%
%
We note that the fraction of sources {\it without} mid-infrared emission is comparatively higher 
compared to the AMGPS masers. This fact, and the intrinsically weaker emission of our
masers, must be carefully considered  in the context of theoretical models for the maser excitation. 

Clearly, the lack of association with mid-infrared emission may also be caused by sensitivity
threshold and/or by optical depth effects. However, since \cite{pandian2011} also
searched for MIPS 24$\mu$m counterparts, they were affected by the same effects and thus the
differences between these two surveys become significant. 
It should also be noted that at present the analysis of the association between mid-infrared 
emission and our 6.7-GHz methanol masers may also be affected by the relatively large beam of the 
Arecibo telescope. In fact, the association rate that we have estimated is based on the 
{\it nominal} position of the Hi-GAL source which, as we have seen in Section~\ref{sec:weak},
may not be precisely coincident with the maser position. In addition, each Hi-GAL clump
may be  generally composed by smaller cores, with different properties, and the core responsible
for the maser emission may not necessarily be coincident with the core associated with the
MIPS 24$\mu$m counterpart.  Therefore, a more 
detailed analysis of the mid-infrared association will require to actually map the masers 
(and, ideally, the Hi-GAL clumps as well) at higher angular resolution in order
to determine their relative positions within the Hi-GAL clump and also with respect to 
the MIPS 24$\mu$m counterpart. However, given all these {\it caveats} it is still quite
surprising that we find the interesting (tentative) trend shown in Fig.~\ref{fig:mips}.

%

\subsection{Correlation between maser intensity and velocity range}
\label{sec:causes}

Our previous discussion indicates that our blind survey of Hi-GAL sources
is more sensitive to the low-flux density methanol masers that have escaped
previous surveys. 
We have also seen in Section~\ref{sec:oh} that theoretical models suggest 
that the presence of a given maser type 
may be linked with the chemical evolution and age of the HMSF region, although this is
difficult to verify observationally without higher angular resolution observations.
However, even our observations have shown a peculiar and interesting correlation. 
In fact, 
Fig.~\ref{fig:DV} shows that the brightest masers tend
to occur in regions with large velocity ranges, and viceversa. 
As discussed in Section~\ref{sec:long}, we can 
tentatively assume that the lack of multiple velocity components toward the 
less bright masers is not an observational effect due to limited 
sensitivity or positional offsets.  
Since maser emission is not isotropic but it is instead 
supposed to be highly beamed (\citealp{alcock1985}, \citealp{elitzur1992}),
then in those regions with relatively few maser spots (and thus with a lower
velocity range), we are less likely to observe any maser emission.
In regions with many maser spots, the probability for the observer
to be aligned with the beaming solid angle of one or more velocity
components is clearly higher. 

Variability of maser emission can be due to both regular and turbulent 
motions of material with different scales and lifetimes, besides
to variations in the pumping source itself. In addition, maser
condensations can split into separate fragments due to interaction
with dense material of the medium where the masers are generated (see, e.g., 
\citealp{lekht2009}). The overall evolution of the maser spots and their
number is not known, but if the number of maser 
spots actually {\it increases} during the evolution of the star forming 
region, then our observations are consistent with these weak 
masers indeed representing an earlier stage.

\section{Conclusions}
\label{sec:conclusions}

We have observed 107 high-mass dust clumps with the Arecibo telescope in search for 
the 6.7-GHz methanol and 6.0-GHz excited OH masers. The clumps were selected from the 
Hi-GAL survey to be relatively massive and visible from Arecibo. 
We detected a total of 32 methanol masers,
with 22 sources being new and weak (median peak flux density 0.07\,Jy) detections,
in the Galactic longitude range $[32^{\circ}.0,59^{\circ}.8]$.

We have compared our results with previous similar surveys, in particular with
the ``Arecibo Methanol Maser Galactic Plane Survey''  \citep{pandian2007a}, and found
that although both source samples approximately cover the same distance range, 
our newly discovered masers are clearly shifted towards much lower integrated flux densities
compared to the AMGPS.  Using 5-pointings cross-scans 
we checked, in a sub-sample of sources, if the masers were being observed off-peak. 
In most cases, the resulting maser peak positions turned out to be very close (i.e., within the Arecibo 
pointing error) to the pointed positions, i.e. the nominal positions of the Hi-GAL sources. 
Thus, most of the methanol masers observed towards our Hi-GAL massive dust clumps appear 
to be intrinsically weaker compared to previously observed masers in unbiased surveys.

The reasons for this lower flux density have yet to be determined, and will likely need
higher-angular resolutions observations. We found no statistically significant correlation
with the physical parameters of the Hi-GAL clumps, except possibly for sources with both maser
types which appear to have higher mass and luminosity compared to sources with just
one type of maser emission. The merged luminosity function of the methanol masers 
detected by us and the AMGPS, shows an essentially flat distribution for luminosities between
$\sim 10^{-7}$ and $\sim 10^{-5}\,L_\odot$ and a relatively quick drop outside of this range.
The intensity of the methanol masers correlates well with the velocity range of the maser emission,
which suggests that the low brightness of these masers is related to the number of maser spots in the
emitting region and their evolution with time.

\begin{acknowledgements}
We wish to thank the staff of the Arecibo Observatory for the support provided
before and during the observations. P.H. acknowledges support from NSF grant AST-0908901.
\end{acknowledgements}

\bibliographystyle{aa} 
\bibliography{refs}    

\clearpage

\appendix


\section{Spectra of methanol masers}
\label{sec:specCH3OH}

\nopagebreak[4]

%
\begin{figure*}[h]
\centering
\includegraphics[width=10.0cm,angle=90]{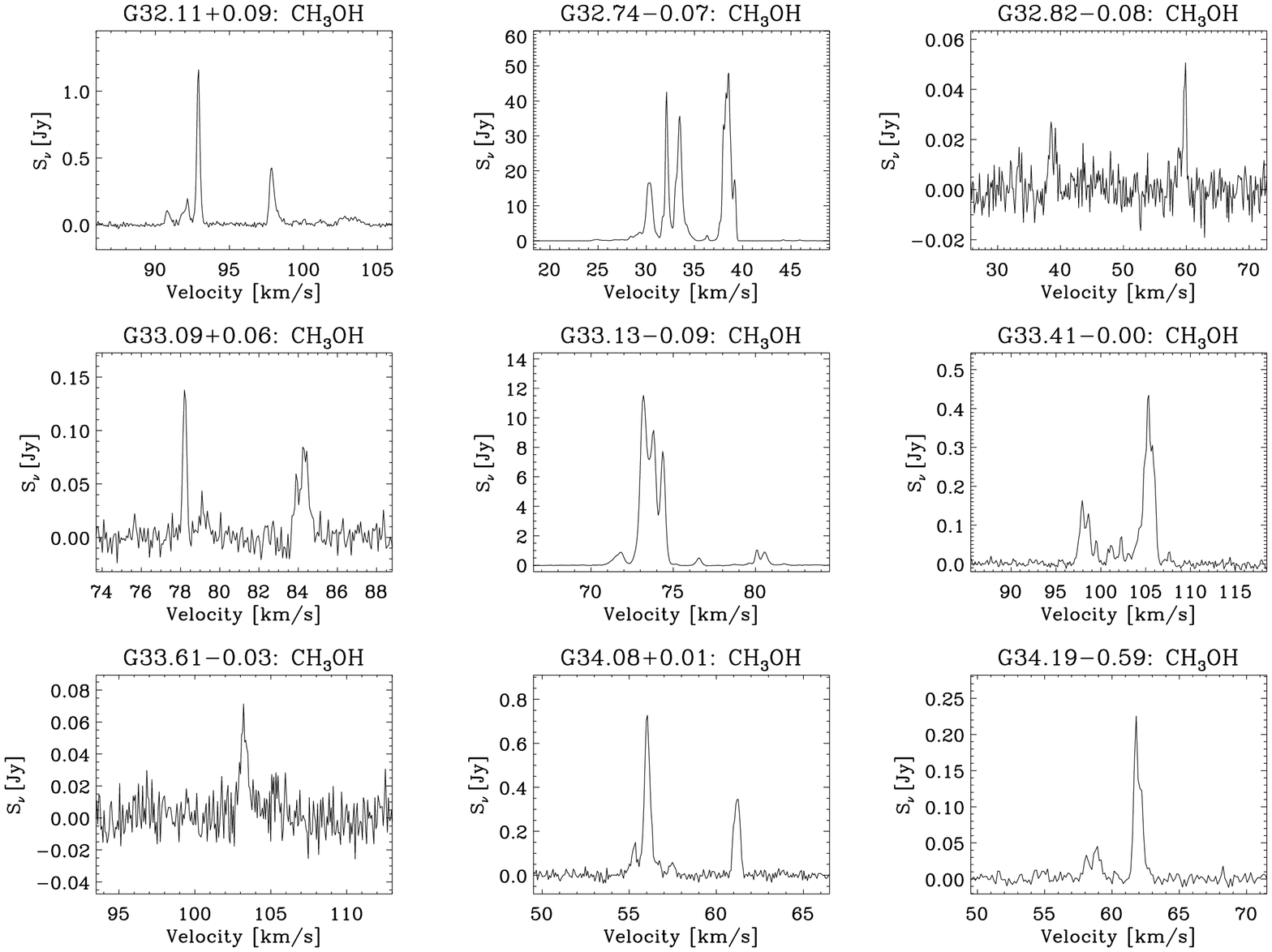}   \\
\includegraphics[width=10.0cm,angle=90]{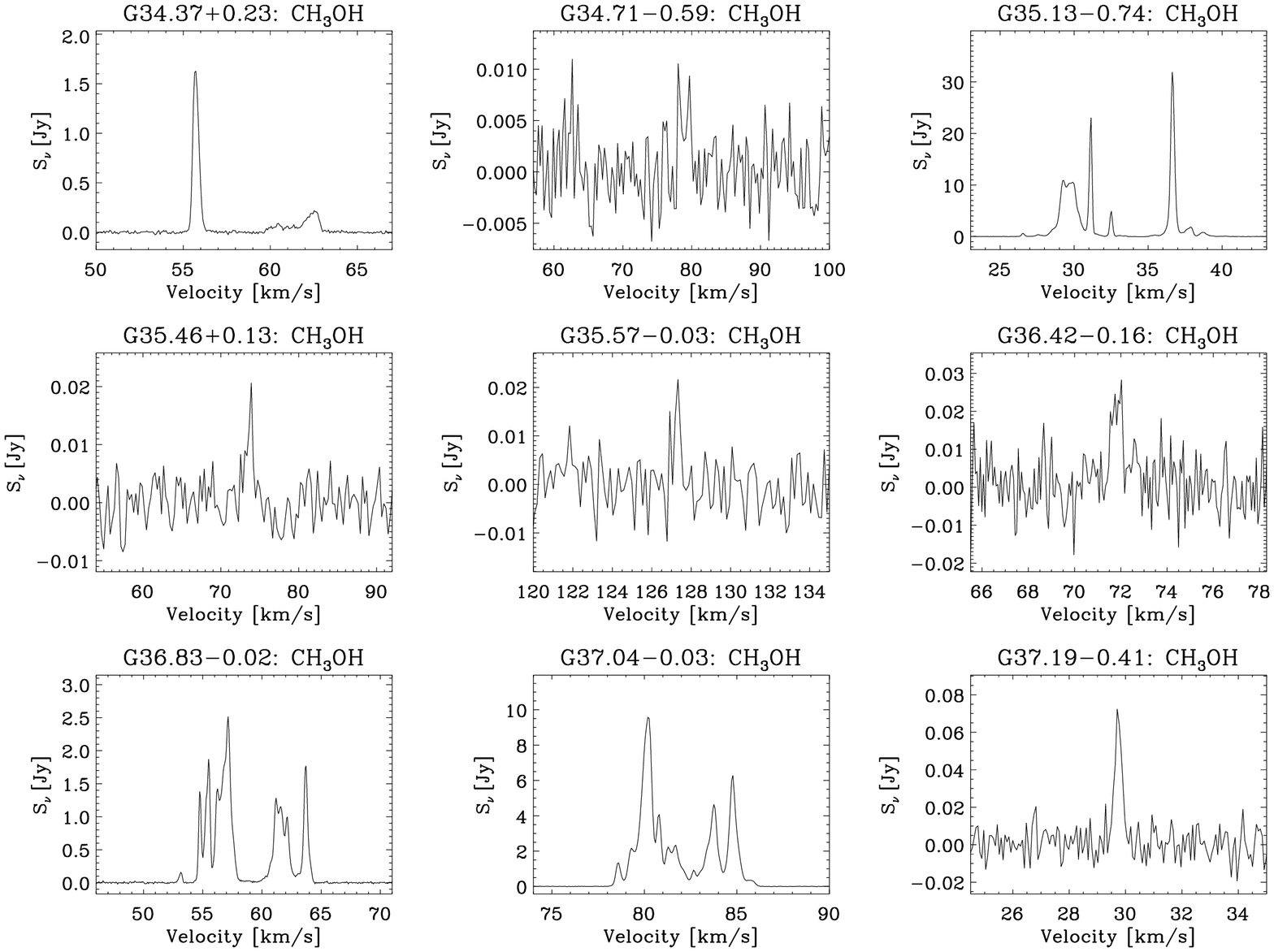}
\caption{
Spectra of methanol masers
  }
\label{fig:spectraCH3OH1}
\end{figure*}

%
\begin{figure*}[h]
\centering
\ContinuedFloat          
\includegraphics[width=10.0cm,angle=90]{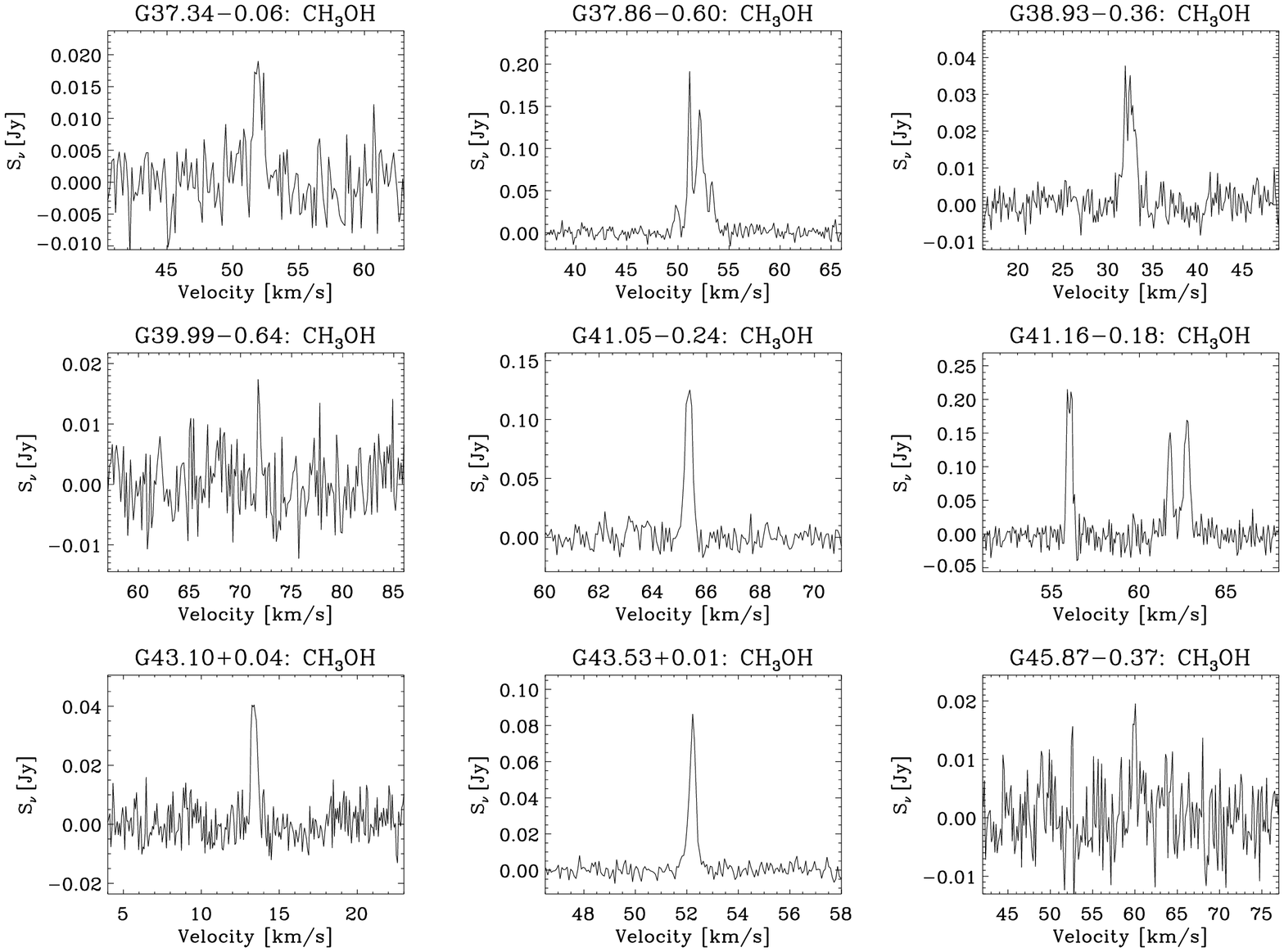}
\includegraphics[width=10.0cm,angle=90]{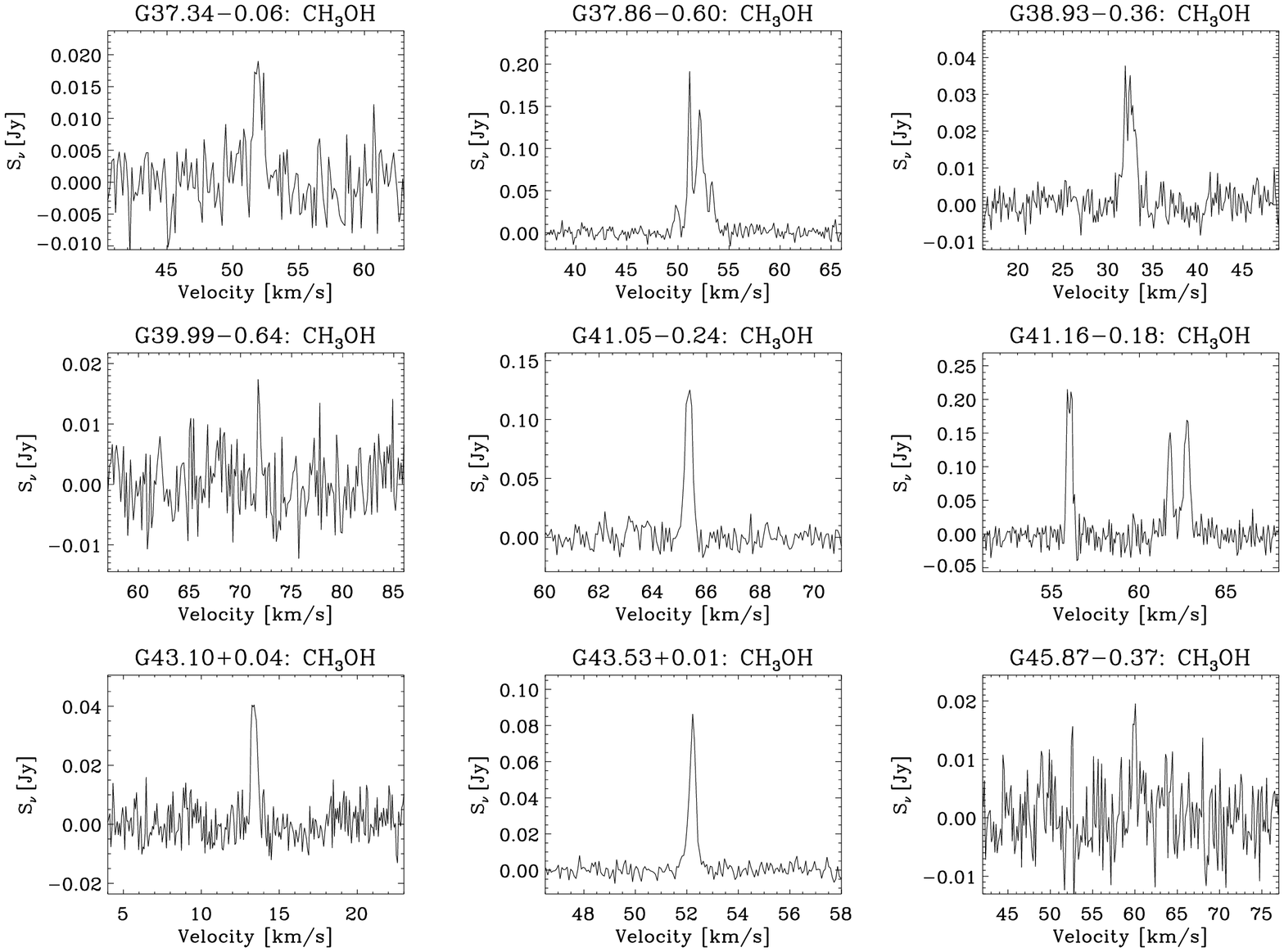}
\caption{
Continued.
  }
\end{figure*}

%
%
%

\clearpage

\section{Spectra of OH masers}
\label{sec:specOH}

\nopagebreak[4]

%
\begin{figure*}[h]
\centering
\includegraphics[width=10.0cm,angle=90]{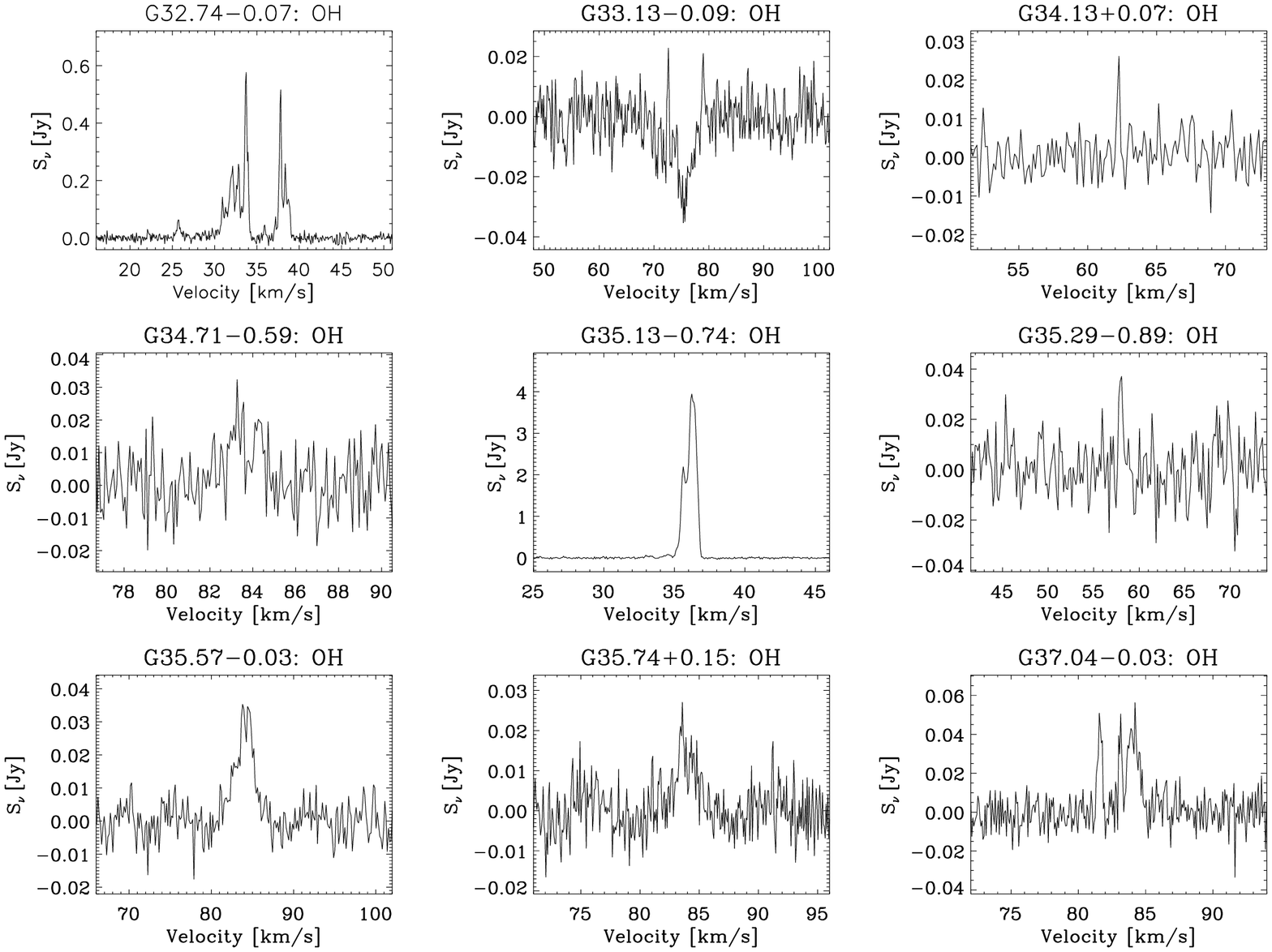}   \\
\includegraphics[width=10.0cm,angle=90]{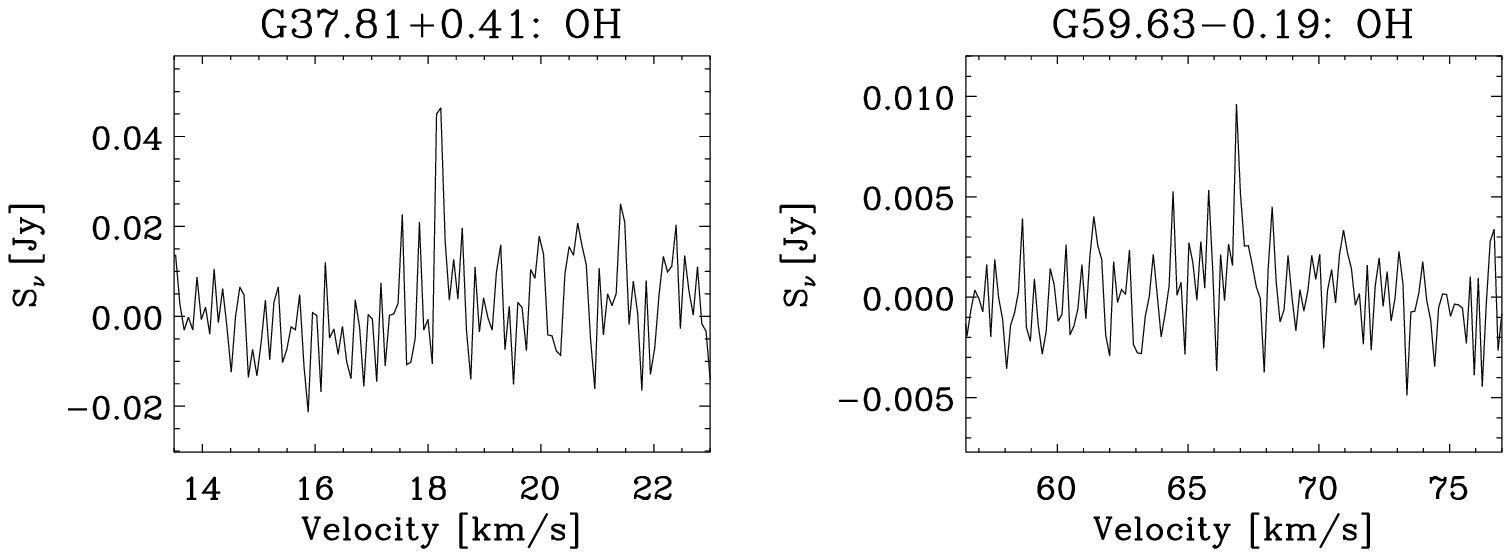}
\caption{
Spectra of OH masers
  }
\label{fig:spectraOH1}
\end{figure*}

%
%
%

\clearpage 

\section{Sources with no methanol detection}
\label{sec:listNODET}

\onecolumn   

%
%
\begin{longtable}{lccr}
\caption{
Hi-GAL sources with no detection of 6.7-GHz methanol masers.
The last column lists the RMS (in mJy) of the final spectrum.
}
\label{tab:listNODET} \\
%
\hline\hline
Name    &  RA            &  DEC       &  RMS  \\                              
        &  [J2000.0]     &  [J2000.0] &  [mJy] \\
\hline
%
G32.45+0.38  & 18:49:11.6  & -00:14:49.0  &    6   \\
G32.45+0.15  & 18:50:00.2  & -00:21:21.5  &   10   \\
G32.03-0.32  & 18:50:58.1  & -00:56:54.8  &   10   \\
G33.11+0.06  & 18:51:33.3  & 00:11:43.1  &   10   \\
G32.98-0.07  & 18:51:47.4  & 00:00:38.2  &    9   \\
G33.26+0.06  & 18:51:48.4  & 00:19:37.9  &   10   \\
G33.70+0.28  & 18:51:50.4  & 00:49:05.9  &   14   \\
G33.71+0.25  & 18:51:57.1  & 00:48:48.6  &    9   \\
G33.49-0.01  & 18:52:30.8  & 00:29:47.6  &   10   \\
G33.02-0.36  & 18:52:54.6  & -00:05:05.9  &   10   \\
G34.46+0.24  & 18:53:20.5  & 01:28:26.0  &    9   \\
G34.13+0.07  & 18:53:21.3  & 01:06:11.2  &    9   \\
G33.33-0.53  & 18:54:03.9  & 00:06:55.4  &    9   \\
G34.00-0.29  & 18:54:26.5  & 00:49:32.8  &   10   \\
G35.42+0.43  & 18:54:27.2  & 02:24:52.7  &    9   \\
G34.94+0.15  & 18:54:33.2  & 01:51:59.0  &    9   \\
G34.69+0.00  & 18:54:39.2  & 01:34:22.0  &    9   \\
G34.24-0.26  & 18:54:46.7  & 01:02:46.0  &    9   \\
G34.93+0.01  & 18:55:01.1  & 01:47:23.5  &    9   \\
G35.56+0.10  & 18:55:53.1  & 02:23:24.5  &    9   \\
G35.60+0.10  & 18:55:56.1  & 02:25:58.3  &    9   \\
G35.74+0.15  & 18:56:01.0  & 02:34:34.0  &    8   \\
G35.44-0.00  & 18:56:03.1  & 02:13:49.1  &   10   \\
G35.61-0.07  & 18:56:36.8  & 02:21:20.9  &    9   \\
G35.68-0.17  & 18:57:04.9  & 02:21:59.0  &    8   \\
G35.52-0.27  & 18:57:08.3  & 02:10:53.9  &    9   \\
G36.40+0.02  & 18:57:42.0  & 03:06:07.9  &    9   \\
G37.49+0.53  & 18:57:53.3  & 04:18:18.7  &    8   \\
G36.45-0.18  & 18:58:31.4  & 03:03:01.5  &    9   \\
G37.17+0.10  & 18:58:49.6  & 03:49:15.0  &    5   \\
G37.61+0.31  & 18:58:51.5  & 04:18:33.6  &   11   \\
G37.81+0.41  & 18:58:53.9  & 04:32:15.1  &    9   \\
G35.29-0.89  & 18:58:57.0  & 01:41:40.0  &   18   \\
G37.42+0.13  & 18:59:09.4  & 04:03:38.0  &    9   \\
G37.37-0.23  & 19:00:23.7  & 03:50:38.9  &    9   \\
G38.19-0.15  & 19:01:35.9  & 04:36:43.9  &    7   \\
G38.42-0.16  & 19:02:04.6  & 04:48:24.6  &    8   \\
G38.32-0.22  & 19:02:06.1  & 04:42:01.8  &    8   \\
G39.25-0.05  & 19:03:12.8  & 05:35:51.2  &   10   \\
G38.69-0.45  & 19:03:35.2  & 04:55:06.8  &   10   \\
G38.92-0.41  & 19:03:52.9  & 05:08:12.9  &    7   \\
G39.49-0.20  & 19:04:10.5  & 05:44:55.0  &    9   \\
G38.35-0.95  & 19:04:44.8  & 04:23:18.9  &    8   \\
G39.85-0.21  & 19:04:53.1  & 06:03:44.5  &   10   \\
G39.26-0.58  & 19:05:07.9  & 05:22:00.0  &    8   \\
G39.36-0.56  & 19:05:13.9  & 05:27:34.1  &    8   \\
G40.36-0.05  & 19:05:15.7  & 06:34:52.3  &    8   \\
G39.99-0.64  & 19:06:39.9  & 05:59:13.7  &    9   \\
G43.23-0.04  & 19:10:33.5  & 09:08:25.0  &    8   \\
G42.15-0.66  & 19:10:45.5  & 07:53:43.8  &    9   \\
G43.51+0.01  & 19:10:51.6  & 09:25:01.3  &    9   \\
G42.23-0.65  & 19:10:53.3  & 07:58:23.6  &    9   \\
G43.30-0.21  & 19:11:16.9  & 09:07:29.7  &    8   \\
G43.32-0.20  & 19:11:17.4  & 09:08:48.8  &    8   \\
G44.48-0.13  & 19:13:12.9  & 10:12:20.9  &   10   \\
G44.49-0.15  & 19:13:17.5  & 10:12:08.8  &   10   \\
G45.95+0.07  & 19:15:14.4  & 11:36:17.6  &    9   \\
G47.04+0.25  & 19:16:41.5  & 12:39:19.9  &   10   \\
G47.09-0.27  & 19:16:42.2  & 12:42:27.0  &    7   \\
G45.38-0.74  & 19:17:07.5  & 10:43:09.6  &   10   \\
G45.88-0.51  & 19:17:14.0  & 11:16:18.9  &    7   \\
G46.42-0.23  & 19:17:16.8  & 11:52:31.0  &    8   \\
G46.17-0.52  & 19:17:49.1  & 11:31:03.7  &    4   \\
G47.00-0.26  & 19:18:28.6  & 12:22:36.8  &    5   \\
G54.45+1.01  & 19:28:25.8  & 19:32:33.5  &    7   \\
G54.39+0.92  & 19:28:38.0  & 19:26:51.9  &    7   \\
G54.11-0.08  & 19:31:48.7  & 18:42:58.0  &    7   \\
G54.22-0.11  & 19:32:10.6  & 18:47:52.9  &    7   \\
G55.74+0.11  & 19:34:27.2  & 20:14:33.1  &    8   \\
G55.15-0.29  & 19:34:45.9  & 19:31:39.4  &    5   \\
G56.06-0.12  & 19:36:00.0  & 20:24:08.7  &    7   \\
G56.89-0.18  & 19:37:58.4  & 21:05:55.6  &    9   \\
\hline
\end{longtable}
%

\end{document}